\documentclass[12pt,a4paper]{article}

\usepackage[utf8]{inputenc}
\usepackage[dvipsnames,svgnames,table]{xcolor}
\usepackage{cite}

\newif{\ifarxiv}
\newif{\ifdraft}
\newif{\ifremarks}

\arxivtrue 
\drafttrue 
\remarkstrue 

\ifdraft\else\fi

\ifremarks\newcommand{\remarktb}[1]{{\renewcommand{\bfdefault}{b}\color[RGB]{0,150,0}{\textbf{Till: #1}}}}\fi
\providecommand{\remarktb}[1]{\ignorespaces}
\ifremarks\newcommand{\comP}[1]{{\color{magenta}{\textbf{Paul: #1}}}}\fi
\providecommand{\comP}[1]{\ignorespaces}
\ifremarks\newcommand{\comT}[1]{{\color[RGB]{0,120,180}{\textbf{Tengis: #1}}}}\fi
\providecommand{\comT}[1]{\ignorespaces}
\ifremarks\newcommand{\comAI}[1]{{\color{blue}{\textbf{AI: #1}}}}\fi
\providecommand{\comAI}[1]{\ignorespaces}
\ifremarks\newcommand{\comClaude}[1]{{\color[RGB]{210,90,0}{\textbf{Claude: #1}}}}\fi
\providecommand{\comClaude}[1]{\ignorespaces}

\usepackage[T1]{fontenc}
\usepackage{lmodern}
\usepackage{microtype}

\usepackage[left=2.3cm,right=2.3cm,top=2cm,bottom=2.5cm]{geometry}
\usepackage{booktabs}
\usepackage{xspace}
\def\etal.{et\penalty50\ al.}
\newcommand*{\eg}{e.\,g.\@\xspace}
\newcommand*{\ie}{i.\,e.\@\xspace}

\makeatletter\newcommand*{\etc}{%
\@ifnextchar{.}%
{etc}%
{etc.\@\xspace}%
}\makeatother

\usepackage[bookmarks=true]{hyperref}

\providecommand{\hypersetup}[1]{}
\providecommand{\hyperref}[2][]{#2}
\providecommand{\texorpdfstring}[2]{#1}
\providecommand{\pdfbookmark}[3][]{}
\newcommand{\email}[1]{\href{mailto:#1}{\nolinkurl{#1}}}

\hypersetup{plainpages=false}
\hypersetup{pdfpagemode=UseOutlines}
\hypersetup{bookmarksnumbered=true}
\hypersetup{bookmarksopen=true}
\hypersetup{pdfstartview=FitH}
\hypersetup{colorlinks=true}
\hypersetup{citecolor=[rgb]{0 .3 0}}
\hypersetup{urlcolor=[rgb]{.35 0 0}}
\hypersetup{linkcolor=[rgb]{0 0 .5}}
\hypersetup{citebordercolor={.6 .9 .6}}
\hypersetup{urlbordercolor={.7 .8 1}}
\hypersetup{linkbordercolor={1 .7 .7}}
\hypersetup{pdfborder={0 0 .5}}

\makeatletter
\let\@myabstract\@empty
\let\@keywords\@empty
\let\@subject\@empty
\providecommand{\affiliation}[1]{\gdef\@affiliation{#1}}
\providecommand{\myabstract}[1]{\gdef\@myabstract{#1}}
\providecommand{\keywords}[1]{\gdef\@keywords{#1}}
\providecommand{\subject}[1]{\gdef\@subject{#1}}
\def\thetitle{\@title}
\def\theauthor{\@author}
\def\theaffiliation{\@affiliation}
\def\theabstract{\@myabstract}
\def\thesubject{\@subject}
\def\thedate{\@date}
\def\thekeywords{\@keywords}
\makeatother
\AtBeginDocument{
\hypersetup{pdftitle={\thetitle}}%
\hypersetup{pdfauthor={\theauthor}}%
\hypersetup{pdfsubject={\thesubject}}%
\hypersetup{pdfkeywords={\thekeywords}}%
}

\usepackage{graphbox}

\usepackage[font=small,labelfont=bf,margin=0.05\textwidth]{caption}

\usepackage{amsmath,amssymb,mathtools}
\usepackage{bbm}
\usepackage[mathbold,autobold,greekcaps,greeklower]{mathfixs}

\newcommand{\nn}{\nonumber}
\allowdisplaybreaks
\numberwithin{equation}{section}

\usepackage{empheq}
\newlength{\widefboxpadding}
\newcommand{\alg}[1]{\mathfrak{#1}}

\RequirePackage[extdef]{delimset}
\makeatletter
\providecommand{\brkleft}[1][r]{\begingroup\def\dlm@use{\delim(.}%
\if r#1 \def\dlm@use{\delim(.}\fi%
\if s#1 \def\dlm@use{\delim[.}\fi%
\if c#1 \def\dlm@use{\delim{.}\fi%
\if a#1 \def\dlm@use{\delim<.}\fi%
\expandafter\endgroup\dlm@use}
\providecommand{\brkright}[1][r]{\begingroup\def\dlm@use{\delim.)}%
\if r#1 \def\dlm@use{\delim.)}\fi%
\if s#1 \def\dlm@use{\delim.]}\fi%
\if c#1 \def\dlm@use{\delim.}}\fi%
\if a#1 \def\dlm@use{\delim.>}\fi%
\expandafter\endgroup\dlm@use}
\makeatother
\newcommand{\matr}[2]{\left(\begin{array}{#1}#2\end{array}\right)}

\DeclareMathOperator{\tr}{tr}

\newcommand{\ii}{i}

\newcommand{\Complex}{\mathbb{C}}

\newcommand{\Ident}{\mathbbm{1}}

\newcommand{\dd}[2][]{\mathinner{d\ifx#1\empty\else{^#1}\fi#2}}

\newcommand{\lO}{\mathcal{O}}

\newcommand{\namedref}[2]{\hyperref[#2]{#1~\ref*{#2}}}
\newcommand{\secref}[1]{\namedref{Section}{#1}}

\newcommand{\tabref}[1]{\namedref{Table}{#1}}

\makeatletter
\def\mr@ignsp#1 {\ifx\:#1\@empty\else #1\expandafter\mr@ignsp\fi}%
\newcommand{\multiref}[1]{\begingroup
\xdef\mr@no@sparg{\expandafter\mr@ignsp#1 \: }%
\def\mr@comma{}%
\@for\mr@refs:=\mr@no@sparg\do{\mr@comma\def\mr@comma{,}\ref{\mr@refs}}%
\endgroup}
\makeatother
\renewcommand{\eqref}[1]{(\multiref{#1})}

\title{Boundary One-Point Functions in the Open XXZ Spin Chain with Generic Integrable Boundaries}

\author{%
Till Bargheer\texorpdfstring{$^{1}$}{},
Tengis Batmunkh\texorpdfstring{$^{1,2}$}{},
Paul Ryan\texorpdfstring{$^{1}$}{}}

\myabstract{We consider the problem of computing observables beyond the spectrum in the open XXZ spin chain with the most generic integrable boundary conditions.
For such boundaries, the residual~$\alg{u}(1)$ symmetry is broken and the familiar Baxter TQ equation describing the spectrum becomes inhomogeneous,
placing the model outside the realm of existing approaches to correlation functions and form factors. We show that the Functional Separation of Variables method,
so far developed mainly for rational, (quasi-)periodic spin chains, is
largely insensitive to these modifications, and works equally well for
diagonal and generic boundaries. Using this method, we compute exact
determinant expressions for a prototypical example of form factors in
open-boundary systems,
namely the longitudinal and transverse boundary-spin one-point functions in arbitrary Hamiltonian eigenstates at finite system size. No constraint relating the two boundary configurations and no reference state are required. We verify the resulting formulae against direct finite-volume diagonalisation. }

\subject{Mathematical Physics}


\begin{document}

\pdfbookmark[1]{Title Page}{title}

\renewcommand{\thefootnote}{\fnsymbol{footnote}}
\setcounter{footnote}{0}

\hfill
\texttt{DESY-26-138}

\mbox{}
\vfill

\begin{center}

{\Large\textbf{\mathversion{bold}\thetitle}\par}

\vspace{1cm}

\textsc{\theauthor}

\bigskip

\begingroup
\footnotesize\itshape

${}^1$\,Deutsches Elektronen-Synchrotron DESY,
Notkestr.~85, 22607 Hamburg, Germany

\smallskip

${}^2$\,Kirchhoff-Institut für Physik, Universität Heidelberg,\\
Im Neuenheimer Feld 227, 69120 Heidelberg, Germany

\endgroup

\bigskip

\begingroup
\small\ttfamily
\email{till.bargheer@desy.de},
\email{tengis.batmunkh@stud.uni-heidelberg.de},
\email{paul.ryan@desy.de}
\endgroup
\par

\vspace{1cm}

\textbf{Abstract}
\vspace{5mm}

\begin{minipage}{12cm}
\theabstract
\end{minipage}

\end{center}

\vfill
\vfill
\newpage

\renewcommand{\thefootnote}{\arabic{footnote}}

\pdfbookmark[1]{\contentsname}{contents}
\setcounter{tocdepth}{2}
\microtypesetup{protrusion=false}
\tableofcontents
\microtypesetup{protrusion=true}


\section{Introduction}
\label{sec:introduction}

The open XXZ spin chain is one of the paradigm examples of a quantum integrable system: an anisotropic nearest-neighbour spin chain which remains exactly solvable in the presence of non-trivial interactions at its two boundaries~\cite{Sklyanin:1988yz,deVega:1993xi,Jimbo:1994ja,Nepomechie:2018dsn,Abetian:2024heu}. For a chain of length $L$, its Hamiltonian takes the form
\begin{equation}
\mathbb{H} = \sum_{\alpha=1}^{L-1}\Big[
\sigma^x_\alpha\sigma^x_{\alpha+1}
+\sigma^y_\alpha\sigma^y_{\alpha+1}
+\cosh\eta\,\sigma^z_\alpha\sigma^z_{\alpha+1}
\Big]
+\mathbb{H}^\partial_-+\mathbb{H}^\partial_+,
\label{eq:hamiltonian}
\end{equation}
where $\sigma^{x,y,z}_\alpha$ are Pauli matrices acting on the $\alpha$-th site, $\eta$ is the anisotropy parameter, and $\mathbb{H}^\partial_\pm$ are the integrable boundary interactions acting on the first and last sites.

A great deal is known about this model: the spectral problem has been studied for several decades and can now be solved~\cite{Sklyanin:1988yz,deVega:1993xi,Cao:2003sbc,Nepomechie:2003vv,Nepomechie:2003ez,Niccoli:2012pi,Faldella:2013qha,Cao:2013bca,Nepomechie:2013ila,Belliard:2013aaa,Belliard:2014rna,Kitanine:2014swa,Maillet:2019hdq,Nepomechie:2019gqt,Nepomechie:2020gwq} for the most general integrable boundary terms $\mathbb{H}^\partial_\pm$, which take the form
\begin{equation}
 \mathbb{H}^\partial_\pm
 =\frac{2\kappa_\pm\sinh\eta}{\sinh\zeta_\pm}
 \Big(e^{\omega_\pm}\sigma^+_{\pm}+e^{-\omega_\pm}\sigma^-_{\pm}\Big)
 +\coth\zeta_\pm\sinh\eta\,\sigma^z_{\pm},
 \label{eqn:bdy-hams}
\end{equation}
where $\sigma^\pm=\tfrac12(\sigma^x\pm\ii\sigma^y)$.
To spell out the notation, in objects such as $\sigma^\pm_\pm$, $\sigma^z_\pm$,
the upper index refers to the component of the Pauli matrices $\sigma^\pm, \sigma^z$,
while the lower index refers to the site on which they act: throughout the paper, a subscript ``$-$'' denotes the first site (site~$1$) and a subscript ``$+$'' denotes the last site (site~$L$) of the chain. Beyond its role as a basic model of quantum integrability, the XXZ
chain also describes real physics: it provides an effective description of quasi-one-dimensional magnetic materials such as
$\mathrm{BaCo}_2\mathrm{V}_2\mathrm{O}_8$~\cite{Takayoshi:2023}.

Despite the tremendous progress on the spectral problem, much less is known about the model's local observables, especially beyond set-ups with the simplest diagonal boundaries where $\kappa_\pm=0$.

\paragraph{Diagonal vs.\ generic boundaries.} For
diagonal boundaries the model can be treated within the standard
framework of the algebraic Bethe ansatz~\cite{Sklyanin:1988yz}. In
this case, the Hamiltonian preserves the $\alg{u}(1)$ symmetry generated by the total spin in the $z$-direction $\mathbb{S}^z$, a pseudovacuum is available, and the spectrum is encoded by a Baxter equation of the form
\begin{equation}
A(u)\,Q(u-\eta)-\tau(u)\,Q(u)+A(-u)\,Q(u+\eta)=0,
\label{eq:introxxz-baxtereq}
\end{equation}
where $\tau(u)$ is the transfer matrix eigenvalue and $Q(u)$ is the corresponding Baxter Q-function. The zeros of $Q(u)$ satisfy ordinary Bethe equations, and the resulting Bethe-state description underlies the determinant scalar-product and quantum inverse-problem techniques used to compute correlation functions~\cite{Kitanine:1998aa,Kitanine:2007openXXZ}.

Generic non-diagonal boundaries change this picture qualitatively. They break the $\alg{u}(1)$ symmetry,
\begin{equation}
[\mathbb{H}^\partial_\pm,\mathbb{S}^z]\neq0,
\end{equation}
so that the standard algebraic Bethe ansatz no longer applies. The spectral problem nevertheless remains exactly solvable through approaches including the off-diagonal Bethe ansatz,
the modified algebraic Bethe ansatz and Separation of
Variables~\cite{Cao:2013qxa,Cao:2013bca,Belliard:2013aaa,
Belliard:2014fsa,Belliard:2014rna,Avan:2015ada,Maillet:2019hdq,
Niccoli:2022xdy}. For completely general integrable boundaries
the spectral data is encoded in an \emph{inhomogeneous} Baxter equation,
\begin{equation}
\mathsf{A}(u)\,Q(u-\eta)-\tau(u)\,Q(u)+\mathsf{A}(-u)\,Q(u+\eta)=-F(u),
\label{eq:introxxz-inhomogeneous-baxtereq}
\end{equation}
where the non-vanishing source term $F(u)$ is fixed by the boundary parameters $\kappa_\pm$, $\zeta_\pm$, and~$\omega_\pm$~\cite{Cao:2013bca,Nepomechie:2013ila,Belliard:2018pvg}.
Here $\mathsf{A}(u)$ denotes the coefficient function for generic boundaries.
We use a different font to distinguish it from the coefficient function $A(u)$ of the diagonal case~\eqref{eq:introxxz-baxtereq}, to which it reduces for $\kappa_\pm\to0$.
Both are given explicitly in \secref{sec:fsov-TQ-diagonal} and \secref{sec:TQ}.

While the source term $F(u)$ in
\eqref{eq:introxxz-inhomogeneous-baxtereq} does not obstruct the
spectral problem, it has more severe consequences for local observables.
The zeros of $Q(u)$ no longer obey the usual factorised Bethe equations,
and the eigenstates are no longer ordinary Bethe states.
Within the modified algebraic Bethe ansatz, they are instead described
by modified Bethe vectors, and determinant representations for their
scalar products have been obtained, first for the twisted XXX chain
\cite{Belliard:2015emq,Belliard:2018hjx,Belliard:2019ebt,Belliard:2019bfz},
and more recently for the XXZ chain with general integrable
boundaries~\cite{Belliard:2021esn}.
These results have not yet been extended to
matrix elements of local operators.

On special lower-dimensional surfaces in the space of boundary parameters, however,
the relevant states can again be described by a homogeneous
Baxter equation, that is $F(u)=0$
\cite{Nepomechie:2003vv,Cao:2003sbc,Nepomechie:2003ez}.
In this constrained setting, Separation of Variables (SoV) has been
successfully employed to obtain local-operator matrix elements and
correlation functions~\cite{Niccoli:2021qhb,Niccoli:2022mfr,Niccoli:2022xdy}.
Away from these constraints, where the Baxter equation remains
genuinely inhomogeneous, a finite-volume computation of physical
local expectation values has not been achieved to date.

\paragraph{Boundary one-point functions.}
In this paper, we address the problem of systematically computing correlation functions in the open XXZ spin chain with the most generic integrable boundaries. We focus on the simplest such physical observable, the boundary one-point functions
\begin{equation}
\frac{\langle\Psi|\sigma^a_-|\Psi\rangle}{\langle\Psi|\Psi\rangle},
\qquad
\frac{\langle\Psi|\sigma^a_+|\Psi\rangle}{\langle\Psi|\Psi\rangle},
\qquad a=z, \pm,
\label{eq:intro-boundary-onepoint}
\end{equation}
which probe the left and right boundaries on sites $1$ and $L$ for eigenstates $\langle \Psi|$ and $|\Psi\rangle$ of the finite chain Hamiltonian.\footnote{For generic complex boundary parameters and anisotropy $\eta$, the left and right Hamiltonian eigenstates $\langle \Psi|$ and $|\Psi\rangle$ are not necessarily Hermitian conjugates. See \secref{sec:openxxz} for precise definitions.} Their longitudinal $z$-components are the boundary magnetisations, while the transverse $\pm$-components measure the polarisation generated by non-diagonal boundary fields. These quantities are the local response observables conjugate to the corresponding boundary couplings, and are sensitive to boundary susceptibilities, boundary bound states and boundary-induced phase transitions~\cite{Alcaraz:1987uk,SkorikKapustin:1995,Fujii:1998qq,Grijalva:2019fkv}. Boundary magnetisations for general boundary configurations have previously been computed through thermodynamic methods. In particular, exact boundary free energies give thermal and ground-state boundary magnetisations in the thermodynamic limit~\cite{Kozlowski:2012fv,Pozsgay:2018ybn}.

\paragraph{Functional Separation of Variables.}
In this paper we compute these quantities in the fully generic configuration. No constraint relating the two boundary matrices is imposed, and the inhomogeneous source term in \eqref{eq:introxxz-inhomogeneous-baxtereq} is retained throughout. The problem considered here is complementary to the thermodynamic computations mentioned above. We seek the one-point functions \eqref{eq:intro-boundary-onepoint} for arbitrary eigenstates at finite $L$, directly in terms of the spectral data (Q-functions). This is the open-chain analogue of the problem of computing finite-volume
expectation values of local operators in arbitrary excited states,
which has been studied extensively for periodic chains and
their field-theory limits~\cite{Pozsgay:2010xd,Pozsgay:2014gza},
in particular within the operatorial SoV framework for the periodic
XXX chain~\cite{Kitanine:2015jna,Niccoli:2020zla}.

Our approach is based on Functional Separation of Variables (FSoV or
Functional SoV). This powerful formalism allows certain correlation functions to be extracted directly from the Baxter equation, without needing to construct the corresponding eigenstates. Following the construction introduced in \cite{Cavaglia:2019pow} and extensively developed in~\cite{Gromov:2020fwh,Gromov:2022waj},\footnote{Boundary overlaps in rational (quasi-)periodic spin chains using the Functional approach have been computed in \cite{Ekhammar:2023iph} and in the operatorial formalism in \cite{Gombor:2021uxz}.} we introduce a functional measure with respect to which the Baxter finite-difference operator is self-adjoint, leading to determinant representations for scalar products and some diagonal form factors. In particular, in this way we are able to compute the boundary one-point functions \eqref{eq:intro-boundary-onepoint} directly in terms of the Baxter Q-function.

A key advantage of the Functional formalism is that, in contrast to
the operatorial formalism, it avoids the necessity to use so-called
Vertex-IRF gauge transformations, which bring one of the two
boundaries to diagonal form at the cost of more complications in the
bulk of the chain. Additionally, we are not required to restrict our
attention to certain boundary configurations where the inhomogeneous
term $F(u)$ in the Baxter equation vanishes
\cite{Niccoli:2022xdy,Niccoli:2022mfr,Niccoli:2026ihg}. As we
will see, the construction works just as easily for general integrable
boundaries as it does for diagonal boundaries. Finally, all of the
computations we perform in the Functional formalism can be done
without needing to introduce local impurities in the spin chain, which
is a crucial feature in the operatorial formalism.

\paragraph{Result.}
We now briefly preview our main result; all objects are defined precisely in the main text.
The central tool is a family of $L$ functional brackets acting on functions of the spectral parameter,
\begin{equation}
 \langle f\rangle_\alpha
 =\oint_{\gamma_+}{\rm d}u\,\mu_\alpha(u)\,f(u)
 +\oint_{\gamma_-}{\rm d}u\,\mu_\alpha(u)\,
 \frac{\mathsf{A}(u+\eta)}{\mathsf{A}(-u)}\,f(u)\,,
 \qquad \alpha=1,\dots,L\,,
\label{eq:intro-brackets}
\end{equation}
where $\gamma_\pm$ are small circles around $u=\pm\eta/2$,
$\mu_\alpha(u)$ are explicit elementary measures given in \eqref{eq:mu-alpha-short},
and $\mathsf{A}(u)$ is the coefficient function of the inhomogeneous Baxter equation
\eqref{eq:introxxz-inhomogeneous-baxtereq} for generic boundaries.
The Baxter finite-difference operator is self-adjoint with respect to each bracket,
and the source term $F(u)$ is annihilated by all of them.
From the Q-function of an eigenstate $|\Psi\rangle$ we build the $L\times L$ matrix
\begin{equation}
 M_{\alpha\beta}=\big\langle Q^2\,p(u)\,s(u)^{\beta-1}\big\rangle_\alpha\,,
\label{eq:intro-M}
\end{equation}
where
\begin{equation}
 p(u)=\sinh^2(2u)-\sinh^2\eta\,,
 \qquad
 s(u)=\sinh(u+\tfrac{\eta}{2})\sinh(u-\tfrac{\eta}{2})\,.
\label{eq:intro-ps}
\end{equation}
The boundary spin operators are generated by the derivative operators
\begin{equation}
 \mathcal D^z_\pm=\partial_{\zeta_\pm}
 +\kappa_\pm\coth\zeta_\pm\,\partial_{\kappa_\pm}\,,
 \qquad
 \mathcal D^\pm_-=\kappa_-\partial_{\kappa_-}\pm\partial_{\omega_-}\,,
 \qquad
 \mathcal D^\pm_+=\kappa_+\partial_{\kappa_+}\pm\partial_{\omega_+}\,,
\label{eq:intro-derivatives}
\end{equation}
acting on the boundary parameters.
Our main result expresses the boundary one-point functions
\eqref{eq:intro-boundary-onepoint} as ratios of determinants,
\begin{equation}
\begin{aligned}
 \big\langle\sigma^z_\pm\big\rangle_\Psi
 &= -\frac{\sinh^2\zeta_\pm}{\mathcal C_1}
 \left[\mathcal D^z_\pm\,\mathcal C'(\eta/2)
 +2\sinh(2\eta)\,\frac{\det M^{(1,\mathcal D^z_\pm)}}{\det M}\right],\\[6pt]
 \big\langle\sigma^{\pm}_-\big\rangle_\Psi
 &= \frac{e^{\mp\omega_-}\sinh\zeta_-}{4\kappa_-\,\mathcal C_1}
 \left[\mathcal D^\pm_-\,\mathcal C'(\eta/2)
 +2\sinh(2\eta)\,\frac{\det M^{(1,\mathcal D^\pm_-)}}{\det M}\right],\\[6pt]
 \big\langle\sigma^{\pm}_+\big\rangle_\Psi
 &= \frac{e^{\mp\omega_+}\sinh\zeta_+}{4\kappa_+\,\mathcal C_1}
 \left[\mathcal D^\pm_+\,\mathcal C'(\eta/2)
 +2\sinh(2\eta)\,\frac{\det M^{(1,\mathcal D^\pm_+)}}{\det M}\right],
\end{aligned}
\label{eq:intro-main-result}
\end{equation}
where $\mathcal C(u)$ is the state-independent part of the transfer matrix eigenvalue,
\begin{equation}
 \tau(u)=\mathcal C(u)+p(u)\sum_{\beta=1}^L H_\beta\, s(u)^{\beta-1}\,,
\label{eq:intro-tau-decomposition}
\end{equation}
with $H_\beta$ the eigenvalues of the $L$ commuting integrals of motion,
and $M^{(1,\mathcal D)}$ is obtained from $M$ by replacing its first column with
$\langle Q\,(\mathcal D Y)\,Q\rangle_\alpha$.
Here $Y$ is the state-independent part of the Baxter equation \eqref{eq:introxxz-inhomogeneous-baxtereq},
so that $\mathcal D Y$ is simply given by
\begin{equation}
   (\mathcal{D} Y)Q(u)=(\mathcal{D} \mathsf{A}(u))\,Q(u-\eta)+(\mathcal{D} \mathsf{A}(-u))\,Q(u+\eta)-(\mathcal{D} \mathcal C(u))\,Q(u)\,.
\end{equation}

Everything on the right-hand side of \eqref{eq:intro-main-result} is determined by the Q-function of the state and explicit functions of the boundary parameters; and so the one-point function is computed without needing to explicitly construct the eigenvectors $\langle \Psi|$ and $|\Psi\rangle$ or differentiate the Q-function with respect to the boundary parameters.
The formula holds for every eigenstate at finite $L$ and for completely generic integrable boundaries.

\paragraph{Paper structure.}
The organisation of the paper is as follows. We first review the
integrable structure and spectral problem of the open XXZ chain with
general boundary matrices, and express the boundary one-point
functions through derivatives of the energy with respect to the
boundary parameters (\secref{sec:openxxz}). Next, we introduce
Functional Separation of Variables directly in the open XXZ setting
(\secref{sec:fsov-correlators}). To make the method transparent, we
first work in the case of diagonal boundaries, where the Baxter equation is homogeneous and
the functional measure, self-adjointness relation and determinant
scalar product can be developed without the source term, and we
complete the boundary one-point-function calculation once in this
simpler setting. Then, we return to
fully generic boundaries in \secref{sec:generic-boundaries}, show how the source term in
\eqref{eq:introxxz-inhomogeneous-baxtereq} is accommodated by the same
functional construction (\secref{sec:fsov-generic-measure}), and
derive the corresponding determinant expressions for the boundary-spin
one-point functions (\secref{sec:fsov-generic-observables}). These
results are tested against direct finite-volume diagonalisation in
\secref{sec:generic-explicit-data}. We conclude in
\secref{sec:conclusions}.

\section{The open XXZ chain and boundary observables}
\label{sec:openxxz}

We now fix our conventions for the open XXZ spin-$1/2$ chain with the most general
integrable boundary conditions, following
\cite{Sklyanin:1988yz,deVega:1993xi,Kitanine:2018sov,Niccoli:2022xdy}. The purpose of this section is
twofold. First, we introduce the transfer matrix and the commuting integrals of motion that
will enter the Functional SoV construction. Second, we show how the boundary-spin one-point functions can be extracted from the derivatives of the eigenvalues of the transfer matrix with respect to the boundary parameters.

\subsection{Integrable structure}
\label{sec:integrable}
\paragraph{R-matrix.}
The bulk integrable structure of the XXZ spin chain is built from the following trigonometric R-matrix
\begin{equation}
R_{12}(u) = \matr{cccc}{
\sinh(u+\eta) & 0 & 0 & 0 \\
0 & \sinh u & \sinh\eta & 0 \\
0 & \sinh\eta & \sinh u & 0 \\
0 & 0 & 0 & \sinh(u+\eta)
},
\label{eq:Rmatrix}
\end{equation}
which acts on $\Complex^2\otimes\Complex^2$. The variable $u$ is the so-called spectral parameter, and $\eta\neq 0$ is a generic fixed complex parameter. The R-matrix satisfies the Yang--Baxter equation (YBE)
\begin{equation}\label{eqn:YBE}
R_{12}(u-v)R_{13}(u-w)R_{23}(v-w)
=
R_{23}(v-w)R_{13}(u-w)R_{12}(u-v)
\end{equation}
on the triple tensor product $\Complex^2\otimes\Complex^2\otimes \mathbb{C}^2$, with the indices $R_{ij}$ denoting that each factor acts non-trivially on the $i$-th and $j$-th copies.

\paragraph{K-matrix.}
For an open chain, compatibility of the boundary reflection with bulk integrability is encoded in
the reflection equation or Boundary Yang--Baxter equation (BYBE)~\cite{Sklyanin:1988yz},
\begin{equation}\label{eqn:BYBE}
R_{12}(u-v)K_1(u)R_{21}(u+v)K_2(v)
=
K_2(v)R_{12}(u+v)K_1(u)R_{21}(u-v),
\end{equation}
whose solutions are called K-matrices. The most general solution is~\cite{deVega:1993xi}\footnote{We restrict our attention to solutions of the reflection equation where the entries of the K-matrices are complex numbers. More general solutions where the entries are operators acting on an appropriate space are relevant for the integrability description of the cusped Wilson loop \cite{Gromov:2021ahm}.}
\begin{equation}
K(u;\zeta,\kappa,\omega)
=
\frac{1}{\sinh\zeta}\matr{cc}{
\sinh(u+\zeta) & \kappa\, e^{\omega}\sinh(2u) \\
\kappa\, e^{-\omega}\sinh(2u) & \sinh(\zeta-u)
}.
\label{eq:Kmatrix}
\end{equation}
Each end of the open chain is described by a K-matrix, which we denote $K_-$ and $K_+$ for the left and right ends respectively
\begin{equation}
K_\pm(u)=K(u\mp\eta/2;\zeta_\pm,\kappa_\pm,\omega_\pm)\,.
\label{eq:crossing}
\end{equation}
Diagonal boundary conditions are recovered when $\kappa_\pm=0$.

\subsection{Transfer matrix and conserved charges}
\label{sec:transfer}

We now put the above ingredients together to build the Hamiltonian and higher conserved charges.
\paragraph{Double-row transfer matrix. }
Following Sklyanin's recipe for boundary integrability, we build the bulk monodromy matrices
\begin{align}\label{eqn:bulk-monodromy}
T_0(u)
&=
R_{01}\big(u-\tfrac{\eta}{2}\big)\cdots
R_{0L}\big(u-\tfrac{\eta}{2}\big),\\
\bar T_0(u)
&=
R_{0L}\big(u-\tfrac{\eta}{2}\big)\cdots
R_{01}\big(u-\tfrac{\eta}{2}\big),
\end{align}
acting on the auxiliary space $V_0\cong\Complex^2$ and the physical Hilbert space
$\mathcal H=(\Complex^2)^{\otimes L}$. From here, we build Sklyanin's double-row monodromy and transfer matrix,
\begin{equation}
U_0(u)=T_0(u)K_{+,0}(u)\bar T_0(u),
\label{eq:doublerow}
\end{equation}
\begin{equation}
t(u)=\tr_0\big[K_{-,0}(u)U_0(u)\big]\in\mathrm{End}(\mathcal H)
\,,
\label{eq:transfer}
\end{equation}
where $K_{\pm,0}$ are the K-matrices acting on the auxiliary space.
As a consequence of the Yang--Baxter equation \eqref{eqn:YBE} and Boundary Yang--Baxter equation \eqref{eqn:BYBE} the transfer matrices mutually commute for any values of the spectral parameters $u$ and $v$
\begin{equation}
[t(u),t(v)]=0,
\end{equation}
so that $t(u)$ generates commuting conserved quantities of the model.

\paragraph{Reference points.}
Two special values of the spectral parameter will be useful in what follows. At
$u=\eta/2$ and $u=\eta/2+\ii\pi/2$ the transfer matrix is proportional to the identity operator on the Hilbert space~\cite{Sklyanin:1988yz,Cao:2013bca,Kitanine:2014swa},
\begin{align}
t(\eta/2)&=\mathcal C_1\,\Ident,
&
\mathcal C_1
&=
\frac{\cosh\eta}{2^{L-1}}
\big[\cosh(2\eta)-1\big]^L,\\
t(\eta/2+\ii\pi/2)&=\mathcal C_2\,\Ident,
&
\mathcal C_2
&=
\frac{(-1)^{L-1}}{2^{L-1}}\cosh\eta\,
\coth\zeta_-\coth\zeta_+
\big[\cosh(2\eta)+1\big]^L.
\label{eq:central-values}
\end{align}
We also need the leading large-$u$ asymptotics of the transfer matrix~\cite{Cao:2013bca,Li:2014gfa,Kitanine:2014swa}
\begin{equation}
t(u)\sim
\mathcal C_3\,\sinh^{2(L+2)}(u)\,\Ident,
\qquad
\mathcal C_3
=
\frac{8\kappa_-\kappa_+\cosh(\omega_--\omega_+)}
{\sinh\zeta_-\sinh\zeta_+}\,,
\label{eq:leading-asymptotics}
\end{equation}
which are state-independent and completely fixed by the boundary parameters. Additionally, the transfer matrix is an even function of $u$, $t(-u)=t(u)$, which can be verified easily starting from the definition \eqref{eq:transfer} together with the YBE \eqref{eqn:YBE} and BYBE \eqref{eqn:BYBE}.

Using these reference values, we can express the transfer matrix in the following convenient form. We will use the shorthand notation
\begin{equation}
     s(u):=\sinh(u+\tfrac{\eta}{2})\sinh(u-\tfrac{\eta}{2})\,.
     \label{eq:S-def}
\end{equation}
Then by interpolation
\begin{equation}
t(u)
=
\mathcal C(u)\,\Ident
+
\big[\sinh^2(2u)-\sinh^2\eta\big]
\sum_{\beta=1}^L\hat H_\beta\, s(u)^{\beta-1},
\label{eq:t-decomposition}
\end{equation}
where
\begin{equation}
\mathcal C(u)
=
\mathcal C_1
+
\frac{\mathcal C_1-\mathcal C_2}{\cosh\eta}
\Big[\sinh^2u-\sinh^2\tfrac{\eta}{2}\Big]
+
\frac{\mathcal C_3}{4}
\big[\sinh^2(2u)-\sinh^2\eta\big]\sinh^{2L}u,
\label{eq:central-full}
\end{equation}
and $\hat H_\beta$, $\beta=1,\ldots,L$, are $L$ mutually commuting non-central
(\ie not proportional to the identity) integrals of motion
\begin{equation}
     [\hat{H}_\beta,\hat{H}_{\beta'}]=0, \quad \beta,\beta'=1,\dots, L\,.
\end{equation}

\paragraph{Hamiltonian from transfer matrix.}

We now relate the physical Hamiltonian \eqref{eq:hamiltonian} to the integrals of motion $\hat{H}_\beta$ coming from the transfer matrix. The precise relation between the Hamiltonian and the transfer matrix is~\cite{Sklyanin:1988yz,deVega:1993xi},
\begin{equation}
\mathbb{H}
=
\frac{\sinh\eta}{\mathcal C_1}
\frac{d}{du}t(u)\Big|_{u=\eta/2}
-C_0,
\qquad
C_0=(L+1)\cosh\eta-\frac{1}{\cosh\eta}.
\label{eq:H-from-logt}
\end{equation}
Hence, in terms of the commuting operators $\hat{H}_\beta$ appearing in \eqref{eq:t-decomposition} we see that the Hamiltonian is simply
\begin{equation}
\mathbb{H}
=
\frac{2\sinh\eta\,\sinh(2\eta)}{\mathcal C_1}\,\hat H_1
+\Big(\frac{\sinh\eta}{\mathcal C_1}\,\mathcal C'(\eta/2)-C_0\Big)\Ident .
\label{eq:H-H1}
\end{equation}
By explicitly performing the computation we indeed verify that the Hamiltonian splits into three terms $\mathbb{H} = \mathbb{H}^\partial_- + \mathbb{H}^B + \mathbb{H}^\partial_+$. The bulk contribution
\begin{equation}
  \mathbb{H}^B = \sum_{\alpha=1}^{L-1}\Big[
\sigma^x_\alpha\sigma^x_{\alpha+1}
+\sigma^y_\alpha\sigma^y_{\alpha+1}
+\cosh\eta\,\sigma^z_\alpha\sigma^z_{\alpha+1}
\Big]
\end{equation}
originates from the R-matrix \eqref{eq:Rmatrix} while the boundary contributions~\eqref{eqn:bdy-hams} are fixed by the K-matrices.
The relation \eqref{eq:H-H1} will be useful in what follows.

\paragraph{Transfer matrix eigenstates.}
For a common eigenstate $|\Psi\rangle$ of the integrals of motion~$\hat{H}_\beta$, we denote the corresponding eigenvalues by $H_\beta$,
so that the transfer matrix eigenvalue $\tau(u)$ with
\begin{equation}\label{eqn:righteigenvec}
     t(u)|\Psi\rangle = \tau(u)|\Psi\rangle
\end{equation}
is obtained from
\eqref{eq:t-decomposition} by $\hat H_\beta\to H_\beta$. In this paper we will also use left eigenvectors $\langle \Psi|$ defined by
\begin{equation}\label{eqn:lefteigenvec}
     \langle \Psi|t(u) = \tau(u)\langle\Psi|\,.
\end{equation}

Since the transfer matrix has non-degenerate spectrum for generic boundary parameters~\cite{Niccoli:2012pi,Faldella:2013qha,Kitanine:2014swa},\footnote{Strictly speaking non-degeneracy of the transfer matrix spectrum has only been established in the presence of local impurities. In rational (quasi-)periodic models non-degeneracy has also been established without impurities. We verified it in our set-up numerically for chains of small length $L=1,2,3,4$ and we assume it to be true in general. We expect that it can be proven using the methods of \cite{mukhin2014spaces,Chernyak:2020lgw,Nepomechie:2019gqt}.} left and right eigenvectors $\langle \Psi_A|$ and $|\Psi_B\rangle$ associated to different transfer matrix eigenvalues $\tau_A(u)$ and $\tau_B(u)$ are orthogonal $\langle \Psi_A|\Psi_B\rangle=0$.
Note that in general a left eigenstate~$\langle \Psi|$ need not be the Hermitian conjugate of the corresponding right eigenstate $|\Psi\rangle$, as we have not imposed any reality structure on the parameters of our model. As such, $\langle\Psi_A|\Psi_B\rangle$ simply denotes the action of a left (row) vector on a right (column) vector.

\paragraph{$\mathfrak{u}(1)$ symmetry breaking.} The R-matrix $R_{12}$ commutes with the diagonal $\mathfrak{u}(1)$ action generated by $\sigma^z$
\begin{equation}
     [R_{12},\sigma^z_1 \otimes 1_2 + 1_1 \otimes \sigma^z_2]=0\,.
\end{equation}
As a result, for diagonal boundary matrices with $\kappa_\pm=0$, this $\mathfrak{u}(1)$ symmetry is inherited by the transfer matrix
\begin{equation}\label{eqn:conserved-Sz}
     [t(u),\mathbb{S}^z]=0,\quad \mathbb{S}^z:=\frac{1}{2}\sum_{n=1}^L \sigma^z_n\,.
\end{equation}
In particular, $\mathbb{S}^z$ commutes with the Hamiltonian $[\mathbb{H},\mathbb{S}^z]=0$.
Thanks to this symmetry, the global spin $\mathbb{S}^z$ is also diagonalised by the transfer matrix eigenstates $|\Psi\rangle$ and $\langle \Psi|$. For generic non-diagonal boundaries $\kappa_\pm\neq 0$, however, this symmetry is broken, and the global $\mathbb{S}^z$ is no longer among the commuting integrals of motion. As stressed in the introduction, this rules out standard methods such as the algebraic Bethe ansatz, which rely on a $\mathfrak{u}(1)$-invariant reference state.

\subsection{Boundary one-point functions}
\label{sec:boundary-magnetization}
\label{sec:boundary-observables}

We now introduce the key quantities we consider in this paper, the boundary one-point functions. It will be convenient to use the following notation. For transfer matrix eigenstates $\langle \Psi|$ and $|\Psi\rangle$ we denote by $\langle \mathcal{O}\rangle_\Psi$ the expectation value of the operator $\lO$
\begin{equation}
\langle\mathcal O\rangle_\Psi
:=
\frac{\langle\Psi|\mathcal O|\Psi\rangle}
{\langle\Psi|\Psi\rangle}.
\label{eqn:vev}
\end{equation}
The longitudinal boundary one-point functions are then defined to be
$\langle\sigma^z_-\rangle_\Psi$ and $\langle\sigma^z_+\rangle_\Psi$, while the transverse boundary one-point functions are $\langle\sigma^\pm_-\rangle_\Psi$ and $\langle\sigma^\pm_+\rangle_\Psi$. For diagonal boundaries, the transverse one-point functions
 vanish because the eigenstates $|\Psi\rangle$ have definite $\mathbb{S}^z$ eigenvalue, as described above, while for generic boundaries they are non-trivial.

\paragraph{Relation to Hamiltonian eigenvalues.}
Now we explain how the boundary one-point functions can be extracted from the spectral data of the model, which will be crucial for computing them via Separation of Variables. By direct computation using \eqref{eqn:bdy-hams} one can verify
\begin{align}
\sigma^z_-
&=
-\frac{\sinh^2\zeta_-}{\sinh\eta}
\Big[\partial_{\zeta_-}
+\kappa_-\coth\zeta_-\,\partial_{\kappa_-}\Big]
\mathbb{H},\nn\\
\sigma^z_+
&=
-\frac{\sinh^2\zeta_+}{\sinh\eta}
\Big[\partial_{\zeta_+}
+\kappa_+\coth\zeta_+\,\partial_{\kappa_+}\Big]
\mathbb{H}\,.
\label{eq:boundary-magnetization}
\end{align}
Similarly, the transverse boundary spin operators are
\begin{align}
\sigma^\pm_-
&=
\frac{e^{\mp\omega_-}\sinh\zeta_-}{4\kappa_-\sinh\eta}
\Big[\kappa_-\partial_{\kappa_-}\pm\partial_{\omega_-}\Big]
\mathbb{H},\nn\\
\sigma^\pm_+
&=
\frac{e^{\mp\omega_+}\sinh\zeta_+}{4\kappa_+\sinh\eta}
\Big[\kappa_+\partial_{\kappa_+}\pm\partial_{\omega_+}\Big]
\mathbb{H}.
\label{eq:boundary-magnetization-pm}
\end{align}
Indeed, the bulk part of $\mathbb{H}$ does not depend on the boundary parameters,
and each derivative combination annihilates all but one
boundary spin operator in $\mathbb{H}^\partial_\pm$.

We now invoke the Hellmann--Feynman theorem of elementary quantum mechanics.
Suppose the conserved charges $\hat{H}_\beta$ depend on a parameter
$\lambda$, and let $|\Psi\rangle$ and $\langle\Psi|$ be right and left
eigenvectors of the transfer matrix for a non-degenerate eigenvalue
$\tau(u)$, with corresponding eigenvalues $H_\beta$ of $\hat H_\beta$.
Non-degeneracy guarantees that $\langle\Psi|\Psi\rangle\neq0$ and that
the eigenvectors can be chosen to depend smoothly on $\lambda$.
Then
\begin{equation}
     \frac{\langle \Psi|\partial_\lambda \hat{H}_\beta|\Psi\rangle}{\langle \Psi|\Psi\rangle}=\partial_\lambda H_\beta\,.
\label{eq:hellmann-feynman}
\end{equation}
Focusing on the Hamiltonian $\mathbb{H}$ and denoting its eigenvalue by $E$, $\mathbb{H}|\Psi\rangle = E\,|\Psi\rangle$,
it then follows from \eqref{eq:boundary-magnetization} and \eqref{eq:boundary-magnetization-pm} that in order to compute the boundary-spin one-point functions
we only need the derivatives of the energy $E$
with respect to the boundary parameters
$\zeta_\pm$, $\kappa_\pm$, and $\omega_\pm$.
For example, from \eqref{eq:boundary-magnetization} we obtain
\begin{equation}
\langle\sigma^z_-\rangle_\Psi
=
-\frac{\sinh^2\zeta_-}{\sinh\eta}
\Big[\partial_{\zeta_-}
+\kappa_-\coth\zeta_-\,\partial_{\kappa_-}\Big]
E,
\label{eq:boundary-HF-example}
\end{equation}
with analogous expressions for the remaining boundary-spin components.

Thus the problem of computing the boundary one-point functions is reduced to determining
the boundary-parameter derivatives $\partial_\lambda E$ of the energy $E$. As we show in the next section, Functional Separation of Variables allows us to easily extract this quantity as
a ratio of determinants built from the Baxter Q-function of the state alone, without needing to construct $|\Psi\rangle$ or vary the boundary parameters to compute the derivatives $\partial_\lambda$.

\section{Functional SoV for diagonal boundaries}
\label{sec:fsov-correlators}

We now begin developing the Functional SoV procedure for boundary one-point functions. Before turning to completely generic boundary fields, we develop the formalism for the open XXZ chain in the simpler setting of diagonal boundaries.

The Functional SoV approach allows one to compute non-trivial form factors and correlators directly from spectral data. Starting from the Baxter equation which encodes the spectrum of conserved quantities, one constructs an appropriate integral operator for which the Baxter equation is self-adjoint. The functional scalar product follows, and from there one can compute a family of diagonal form factors.

We start by reviewing the central ingredient, the Baxter equation.

\subsection{Baxter equation for diagonal boundaries}
\label{sec:fsov-TQ-diagonal}

As a result of the restriction to diagonal boundaries, the global $\mathbb{S}^z$ in \eqref{eqn:conserved-Sz} is a conserved quantity. We label its eigenvalues as $L/2-M$, where $L$ is the length of the chain and $0\leq M\leq L$ is the number of spin flips.

The spectrum of the transfer matrix \eqref{eq:transfer}
can be encoded in a second-order finite-difference Baxter equation, see for instance~\cite{Sklyanin:1988yz,Kitanine:2018sov,Maillet:2019hdq}. An eigenvalue $\tau(u)$ is
associated with an even trigonometric polynomial
\begin{equation}
 Q(u)=\prod_{j=1}^{M}\big(\sinh^2u-\sinh^2u_j\big)\,,
 \label{eq:Qfn-diagonal}
\end{equation}
where $u_j$ are the so-called Bethe roots, familiar from Bethe ansatz approaches \cite{Sklyanin:1988yz}. The Baxter equation relates $\tau(u)$ and $Q(u)$ and reads
\begin{equation}
 A(u)Q(u-\eta)-\tau(u)Q(u)+A(-u)Q(u+\eta)=0,
 \label{eq:TQ-diagonal}
\end{equation}
with
\begin{equation}
 A(u)=(-1)^L
 \frac{\sinh(u+\zeta_--\tfrac\eta2)\,\sinh(u+\zeta_+-\tfrac\eta2)\,\sinh(2u+\eta)}
 {\sinh\zeta_-\,\sinh\zeta_+\,\sinh(2u)}\,a(u)d(-u),
 \label{eq:A-diagonal}
\end{equation}
and
\begin{equation}
 a(u)=\sinh^L\big(u+\tfrac\eta2\big),\qquad d(u)=\sinh^L\big(u-\tfrac\eta2\big).
 \label{eq:ad-bulk}
\end{equation}

Using the Baxter equation, the spectrum of all conserved charges $\hat{H}_\beta$ entering the transfer matrix \eqref{eq:t-decomposition} can be extracted. In particular, one can extract the eigenvalue $E$ of the Hamiltonian $\mathbb{H}$ as
\begin{equation}
 E=\frac{\sinh\eta}{\mathcal C_1}A'(\eta/2)-C_0-2\sinh\eta\,\frac{Q'(\eta/2)}{Q(\eta/2)}\,.
\end{equation}
It has been established rigorously for the set-up with local impurities that there is a one-to-one correspondence between solutions of the Baxter equation and physical eigenvalues~$\tau(u)$ of the transfer matrix~$t(u)$~\cite{Niccoli:2012pi,Faldella:2013qha,Kitanine:2014swa,Kitanine:2018sov}, \ie that for every physical eigenvector and eigenvalue there is a corresponding Q-function. This result should extend to our set-up by analyticity, and we verified it for chains of small length $L=1,2,3,4$. We will assume that this and non-degeneracy of the spectrum are true in general.
Note that for explicit computations, \eg when solving the Baxter equation numerically, it is often convenient to regard $Q$ directly as a polynomial in
$\sinh^2u$, for example
\begin{equation}
 Q(u)=\sum_{j=0}^{M}c_j\sinh^{2j}u,\qquad c_M=1\,,
 \label{eq:Qfn-diagonal-poly}
\end{equation}
rather than solving explicitly for its roots.

\subsection{Functional measure and self-adjointness}
\label{sec:fsov-measure}

With the Baxter equation reviewed, we will now complete the next step
in the Functional SoV procedure, which is to construct an appropriate
measure for which it is a self-adjoint operator on a suitable space of functions.

\paragraph{Baxter equation as a difference operator.}
The starting point is to introduce the shift operator $Df(u)=f(u+\eta/2)$ and write
\eqref{eq:TQ-diagonal} as
\begin{equation}
 \mathcal O=A(u)D^{-2}+A(-u)D^2-\tau(u),\qquad \mathcal OQ=0.
\end{equation}
Note that there is a corresponding operator $\mathcal O$ for each transfer matrix eigenvalue $\tau(u)$.

Following \cite{Cavaglia:2019pow}, we now seek linear forms
$\langle f\rangle$, defined on generic entire functions $f(u)$,
with respect to which the finite-difference operator $\mathcal{O}$
is self-adjoint,
$\langle f\,\mathcal Og\rangle=\langle g\,\mathcal Of\rangle$. We make an ansatz for $\langle f\rangle$ of the following form
\begin{equation}
  \langle f \rangle = \oint_{\gamma_+}{\rm d}u\,\mu_+(u)f(u)+\oint_{\gamma_-}{\rm d}u\,\mu_-(u)f(u)\,,
  \label{eq:bracket-ansatz-short}
\end{equation}
where the measures $\mu_\pm(u)$ are assumed to have poles at $\pm\eta/2$, and $\gamma_+$ is a small counterclockwise circle around $\eta/2$ and $\gamma_-=\gamma_+-\eta$ is a small circle around $-\eta/2$.

\paragraph{Consistency conditions on the measure.}
Let us now derive conditions on $\mu_\pm$ by demanding that
$\langle f\mathcal{O} g\rangle = \langle g \mathcal{O}f\rangle$.
Writing out the action of $\mathcal O$ in \eqref{eq:bracket-ansatz-short}
we find
\begin{equation}
  \begin{split}
  \langle f \mathcal{O} g\rangle
  & = \oint_{\gamma_+} {\rm d}u\,\mu_+(u) f(u)A(u)g(u-\eta)
    +\oint_{\gamma_+} {\rm d}u\,\mu_+(u) f(u)A(-u)g(u+\eta) \\
  & \quad -\oint_{\gamma_+} {\rm d}u\,\mu_+(u) f(u)\tau(u)g(u) \\
  & \quad +\oint_{\gamma_-} {\rm d}u\,\mu_-(u) f(u)A(u)g(u-\eta)
    +\oint_{\gamma_-} {\rm d}u\,\mu_-(u) f(u)A(-u)g(u+\eta) \\
  & \quad -\oint_{\gamma_-} {\rm d}u\,\mu_-(u) f(u)\tau(u)g(u)\,.
  \end{split}
  \label{eq:fOg-expanded}
\end{equation}
The two terms containing $\tau(u)$ are manifestly symmetric
under $f\leftrightarrow g$, so we only need to consider the
four terms containing $A$,
\begin{equation}
  \begin{split}
  &\oint_{\gamma_+} {\rm d}u\,\mu_+(u)A(u)\,f(u)\,g(u-\eta)
  +\oint_{\gamma_+} {\rm d}u\,\mu_+(u)A(-u)\,f(u)\,g(u+\eta) \\
  &+\oint_{\gamma_-} {\rm d}u\,\mu_-(u)A(u)\,f(u)\,g(u-\eta)
  +\oint_{\gamma_-} {\rm d}u\,\mu_-(u)A(-u)\,f(u)\,g(u+\eta)\,.
  \end{split}
 \label{eq:four-terms}
\end{equation}
In the first integral we substitute $u\to u+\eta$,
which maps $\gamma_+$ onto $\gamma_-$.
It then becomes
\begin{equation}
 \oint_{\gamma_-}{\rm d}u\,\mu_+(u+\eta)A(u+\eta)\,f(u+\eta)\,g(u)\,.
\end{equation}
This has the same form as the fourth integral in
\eqref{eq:four-terms} with the roles of $f$ and $g$ exchanged,
so the two combine into an expression
symmetric under $f\leftrightarrow g$ provided that
\begin{equation}
 \mu_-(u)\,A(-u)=\mu_+(u+\eta)\,A(u+\eta)
 \label{eq:measure-eq-short}
\end{equation}
near $u=-\eta/2$.
On the other hand, the second and third integrals in \eqref{eq:four-terms}
cannot be paired in this way, so we demand that they vanish
for all entire $f$ and $g$.
This requires that
\begin{equation}
 \mu_+(u)\,A(-u)\ \text{is regular at}\ u=\tfrac{\eta}{2}\,,
 \label{eq:regularity-short-plus}
\end{equation}
\begin{equation}
 \mu_-(u)\,A(u)\ \text{is regular at}\ u=-\tfrac{\eta}{2}\,.
 \label{eq:regularity-short}
\end{equation}
From the explicit form \eqref{eq:A-diagonal} of $A(u)$,
the factor $a(u)d(-u)$ has a zero of order $2L$ at $u=-\eta/2$,
and the factor $\sinh(2u+\eta)$ has a simple zero there,
so $A(u)$ vanishes to order $2L+1$ at $u=-\eta/2$
and likewise $A(-u)$ vanishes to order $2L+1$ at $u=\eta/2$.
Hence $\mu_+(u)$ may have a pole of order at most $2L+1$ at $u=\eta/2$,
and $\mu_-(u)$ a pole of order at most $2L+1$ at $u=-\eta/2$.

Under these conditions the remaining terms equal
\begin{equation}
 \oint_{\gamma_-}{\rm d}u\,\mu_-(u)\,A(-u)\,
 \big[f(u+\eta)\,g(u)+f(u)\,g(u+\eta)\big]\,,
\end{equation}
which is symmetric under $f\leftrightarrow g$,
so $\mathcal{O}$ is self-adjoint.

\paragraph{Solving the measure equation.}
It is now easy to find solutions of the measure
equation~\eqref{eq:measure-eq-short} obeying the regularity
conditions~\eqref{eq:regularity-short-plus} and~\eqref{eq:regularity-short}.

We construct a particular solution by introducing an $\eta$-periodic function $\mu(u)$, meromorphic in
neighbourhoods of $u=\pm\eta/2$, with a pole of order at most $2L+1$
at $u=\eta/2$ and hence also at $u=-\eta/2$.
We then set
\begin{equation}
 \mu_+(u)=\mu(u)\,,
\end{equation}
\begin{equation}
 \mu_-(u)=\mu(u)\,\frac{A(u+\eta)}{A(-u)}\,,
\end{equation}
so that \eqref{eq:measure-eq-short} holds by periodicity of $\mu$,
and the bracket \eqref{eq:bracket-ansatz-short} reads
\begin{equation}
 \langle f\rangle=\oint_{\gamma_+}{\rm d}u\,\mu(u)\,f(u)
 +\oint_{\gamma_-}{\rm d}u\,\mu(u)\,\frac{A(u+\eta)}{A(-u)}\,f(u)\,.
 \label{eq:bracket-two-contours}
\end{equation}
The ratio $A(u+\eta)/A(-u)$ is regular and non-zero at $u=-\eta/2$,
where both numerator and denominator equal $A(\eta/2)$.
The regularity conditions \eqref{eq:regularity-short-plus}
and \eqref{eq:regularity-short} hold
since $A(-u)$ has a zero of order $2L+1$ at $u=\eta/2$
and $A(u)$ has a zero of order $2L+1$ at $u=-\eta/2$,
which cancel the poles of $\mu$.

\paragraph{Family of brackets.}
Since we can choose any $\mu$ satisfying the above requirements, we obtain a whole family of brackets.
We select $L$ of them by choosing the $\eta$-periodic measures
\begin{equation}
 \mu_\alpha(u):=\frac{1}{\big(1+e^{2\pi\ii u/\eta}\big)^{2L+3-2\alpha}}\,,
 \qquad \alpha=1,\dots,L\,.
 \label{eq:mu-alpha-short}
\end{equation}
The corresponding brackets are denoted $\langle f\rangle_\alpha$ and are given by
\eqref{eq:bracket-two-contours} with $\mu$ replaced by $\mu_\alpha$,
\begin{equation}
 \langle f\rangle_\alpha=\oint_{\gamma_+}{\rm d}u\,\mu_\alpha(u)\,f(u)
 +\oint_{\gamma_-}{\rm d}u\,\mu_\alpha(u)\,\frac{A(u+\eta)}{A(-u)}\,f(u)\,,
 \qquad \alpha=1,\dots,L\,.
 \label{eq:bracket-alpha-diag}
\end{equation}
Our choice of periodic functions in \eqref{eq:mu-alpha-short} is
essentially canonical. Any admissible $\mu$ has a pole of order at
most $2L+1$ at $u=\eta/2$, so its bracket is a linear combination of
the brackets with $\mu=(1+e^{2\pi\ii u/\eta})^{-k}$, $k=1,\dots,2L+1$.
On even functions, the brackets with even $k$ reduce to those with
lower odd $k$, and the bracket with $k=1$ only probes $f(\eta/2)$,
where all functions entering the constructions below vanish. Any other
set of $L$ admissible measures therefore changes the rows of $M$ and
$y^{(\lambda)}$ by a common linear map, which is invertible as long as
the resulting $M$ remains non-degenerate. This rescales $\det M$ by a
state-independent constant but leaves the ratios in Cramer's rule, and
hence all physical quantities, unchanged, see also~\cite{Cavaglia:2021mft}.
We will make use of all $L$ of these brackets in what follows.

\subsection{Functional scalar product and diagonal form factors}
\label{sec:fsov-orthogonality}

\paragraph{Functional scalar product.}
Armed with the brackets $\langle -\rangle_\alpha$ we can now construct the Functional SoV scalar product, following the approach of \cite{Cavaglia:2019pow}. Let $Q_A,Q_B$ be the Q-functions of two transfer matrix eigenvalues $\tau_A(u)$ and $\tau_B(u)$, satisfying
$\mathcal O_AQ_A=0=\mathcal O_BQ_B$.  Self-adjointness gives
\begin{equation}
 \big\langle Q_B(\mathcal O_A-\mathcal O_B)Q_A\big\rangle_\alpha=0.
 \label{eq:fsov-diag-orthog-step}
\end{equation}
Using the decomposition \eqref{eq:t-decomposition}, whose central piece is
state-independent, this becomes
\begin{equation}
 \sum_{\beta=1}^L\big(H_\beta^A-H_\beta^B\big)M^{AB}_{\alpha\beta}=0,
 \qquad
 M^{AB}_{\alpha\beta}:=
 \Big\langle Q_AQ_B\, p(u)\, s(u)^{\beta-1}\Big\rangle_\alpha,
 \label{eq:diag-orthogonality-tau-exp}
\end{equation}
where $s(u)$ is defined in \eqref{eq:S-def} and
\begin{equation}
  p(u):=\sinh^2(2u)-\sinh^2\eta.
 \label{eq:fsov-P-def}
\end{equation}
For distinct eigenstates the vector of integral-of-motion differences is non-zero, hence the corresponding determinant of the linear system must vanish
\begin{equation}
 \det M^{AB} :=\det_{1\leq\alpha,\beta\leq L}M^{AB}_{\alpha\beta}\propto\delta_{AB}\,.
 \label{eq:diag-scalarproduct}
\end{equation}
This is the functional scalar product\footnote{A similar scalar product was constructed in rational open spin chains in \cite{Gromov:unpublished}.} which manifestly vanishes for two distinct states. For $A=B$ we assume
that $\det M^{AA}\neq0$ for every eigenstate.
We verified this numerically for chains of length $L=1,2,3,4$,
both for diagonal and for generic boundary parameters.
We expect that it can be proven by introducing local impurities
on the chain and relating $\det M^{AB}$ to the scalar product
of the operatorial SoV approach~\cite{Kitanine:2018sov},
as was done for periodic chains in \cite{Gromov:2019wmz,Gromov:2020fwh}.
We will not present such a proof here. Instead, numerical evidence
is presented in \secref{sec:generic-explicit-data}.

\paragraph{Diagonal form factors.}

We can now use the developed machinery to compute diagonal form factors, following the approach of~\cite{Cavaglia:2018lxi,Cavaglia:2019pow}.

Let $\lambda$ be any parameter of the model, which we assume to be one of the boundary parameters. All of the functions entering the Baxter operator $\mathcal{O}$, the Q-functions, as well as the integral of motion eigenvalues $H_\beta$ are assumed to depend on this parameter. As a consequence of the self-adjointness property of the brackets $\langle f\rangle_\alpha$, one can easily show
\begin{equation}
     \langle Q (\partial_\lambda \mathcal{O})Q\rangle_\alpha=0,\quad \alpha=1,\dots,L\,,
\label{eqn:diff-baxter}\end{equation}
where $Q$ is the Q-function corresponding to some fixed eigenstate $|\Psi\rangle$ of the transfer matrix. Note that we have dropped the state labels $A$ of the previous subsection, as now we only refer to a single fixed state.

We now expand the difference operator $\mathcal{O}$ into a state-independent part $Y$ and a part depending on the integrals of motion $H_\beta$
\begin{equation}
 \mathcal O= Y- p(u)\sum_{\beta=1}^LH_\beta\, s(u)^{\beta-1},
 \qquad
 Y:=A(u)D^{-2}+A(-u)D^2-\mathcal C(u)
 \label{eq:fsov-diag-Y}
\end{equation}
where we recall that $A(u)$, $\mathcal{C}(u)$, $p(u)$ and $s(u)$ are defined in \eqref{eq:A-diagonal}, \eqref{eq:central-full}, \eqref{eq:fsov-P-def} and \eqref{eq:S-def}, respectively. Hence, plugging into \eqref{eqn:diff-baxter}, we obtain
\begin{equation}
     \langle Q (\partial_\lambda Y)Q \rangle_\alpha - \sum_{\beta=1}^L\langle Q^2 p(u) s(u)^{\beta-1}\rangle_\alpha \,\partial_\lambda H_{\beta} = 0,  \quad \alpha=1,\dots,L\,.
\end{equation}

Thus the derivatives of the integrals of motion $\partial_\lambda H_\beta$ obey the $L\times L$ linear system
\begin{equation}
 \sum_{\beta=1}^L M_{\alpha\beta}\,\partial_\lambda H_\beta=y^{(\lambda)}_\alpha,
 \qquad
 M_{\alpha\beta}=\Big\langle Q^2 p(u)\, s(u)^{\beta-1}\Big\rangle_\alpha,
 \qquad
 y^{(\lambda)}_\alpha=\big\langle Q(\partial_\lambda\,Y)Q\big\rangle_\alpha.
 \label{eq:fsov-diag-derivative-system}
\end{equation}
Cramer's rule then gives
\begin{equation}
 \partial_\lambda H_\beta=\frac{\det M^{(\beta,\lambda)}}{\det M},
 \label{eq:diag-formfactor}
\end{equation}
where $M^{(\beta,\lambda)}$ is obtained from $M$ by replacing column $\beta$ with
$y^{(\lambda)}$. Note that, crucially, the denominator is non-zero
by the assumption $\det M^{AA}\neq0$ stated above. By the Hellmann--Feynman theorem \eqref{eq:hellmann-feynman},
this is directly the diagonal matrix element
\begin{equation}
 \partial_\lambda H_\beta=
 \frac{\langle\Psi|\partial_\lambda\hat H_\beta|\Psi\rangle}{\langle\Psi|\Psi\rangle}.
\end{equation}
The important conceptual point is that the derivative $\partial_\lambda H_\beta$ does not require us to differentiate the Q-function in $\lambda$, which can be rather involved, especially at large system size where we may only know the Q-function numerically.

We now finally compute the boundary one-point functions as an explicit example of this construction.

\subsection{Boundary magnetisation}
\label{sec:fsov-formfactors}

As we described in Section~\ref{sec:boundary-magnetization} the boundary spin operators are obtained from
the Hamiltonian as in \eqref{eq:boundary-magnetization},
\begin{equation}
 \sigma^z_-=-\frac{\sinh^2\zeta_-}{\sinh\eta}
 \Big[\partial_{\zeta_-}+\kappa_-\coth\zeta_-\partial_{\kappa_-}\Big]
 \mathbb{H},
 \label{eq:fsov-sigmaz-def}
\end{equation}
and similarly
\begin{equation}
 \sigma^z_+=-\frac{\sinh^2\zeta_+}{\sinh\eta}
 \Big[\partial_{\zeta_+}+\kappa_+\coth\zeta_+\partial_{\kappa_+}\Big]
 \mathbb{H}.
 \label{eq:fsov-sigmaz-def-right}
\end{equation}
For diagonal boundaries we have $\kappa_\pm=0$, and the transverse one-point functions $\langle\sigma^\pm_-\rangle_\Psi$ and $\langle\sigma^\pm_+\rangle_\Psi$ vanish by
$\alg{u}(1)$ symmetry, so we will not consider them here.

By \eqref{eq:H-H1}, only $H_1$ enters $\mathbb{H}$. Specialising \eqref{eq:fsov-sigmaz-def} to diagonal boundaries and using
\eqref{eq:diag-formfactor} with $\beta=1$ therefore yields a single ratio of
determinants,
\begin{equation}
 \langle\sigma^z_-\rangle_\Psi=
 -\frac{\sinh^2\zeta_-}{\mathcal C_1}
 \left[
 \partial_{\zeta_-}\mathcal C'(\eta/2)
 +2\sinh(2\eta)\,\frac{\det M^{(1,\zeta_-)}}{\det M}
 \right]\,.
 \label{eq:fsov-boundarymag-compact}
\end{equation}
Here $M^{(1,\zeta_-)}$ is $M$ with its first column replaced by
$y^{(\zeta_-)}_\alpha=\big\langle Q(\partial_{\zeta_-}Y)Q\big\rangle_\alpha$.
The right-boundary result follows by $\zeta_-\to\zeta_+$ and the corresponding boundary
derivative. This expression holds for all eigenstates $|\Psi\rangle$ of the finite-volume spin chain.

\paragraph{Explicit example.}
As an illustration, consider $L=2$.
The two matrices entering \eqref{eq:fsov-boundarymag-compact} are
\begin{equation}
 M=\begin{pmatrix}
 \langle Q^2\, p\rangle_1 & \langle Q^2\, p\, s\rangle_1\\[4pt]
 \langle Q^2\, p\rangle_2 & \langle Q^2\, p\, s\rangle_2
 \end{pmatrix},
 \qquad
 M^{(1,\zeta_-)}=\begin{pmatrix}
 \langle Q\,(\partial_{\zeta_-}Y)\,Q\rangle_1 & \langle Q^2\, p\, s\rangle_1\\[4pt]
 \langle Q\,(\partial_{\zeta_-}Y)\,Q\rangle_2 & \langle Q^2\, p\, s\rangle_2
 \end{pmatrix},
 \label{eq:fsov-L2-matrices}
\end{equation}
with
\begin{equation}
 \big(\partial_{\zeta_-}Y\big)Q
 =\partial_{\zeta_-}A(u)\,Q(u-\eta)+\partial_{\zeta_-}A(-u)\,Q(u+\eta)
 -\partial_{\zeta_-}\mathcal C(u)\,Q(u)\,.
\end{equation}
The state-independent term in \eqref{eq:fsov-boundarymag-compact}
only requires the central function \eqref{eq:central-full}.
For diagonal boundaries $\mathcal C_3=0$, since it is proportional to $\kappa_-\kappa_+$,
and for $L=2$ the remaining constants \eqref{eq:central-values} reduce to
\begin{equation}
 \mathcal C_1=2\cosh\eta\sinh^4\eta,
 \qquad
 \mathcal C_2=-2\cosh^5\eta\,\coth\zeta_-\coth\zeta_+\,.
\end{equation}
Evaluating $\partial_{\zeta_-}\mathcal C'(\eta/2)$ with these values,
\eqref{eq:fsov-boundarymag-compact} becomes
\begin{equation}
 \langle\sigma^z_-\rangle_\Psi=\coth^3\eta\,\coth\zeta_+
 -\frac{2\sinh^2\zeta_-}{\sinh^3\eta}\,
 \frac{\det M^{(1,\zeta_-)}}{\det M}\,.
 \label{eq:fsov-L2-boundarymag}
\end{equation}

To summarise, we have managed to compute the boundary one-point functions directly from spectral data for the open XXZ chain with diagonal boundaries, \ie the Q-function solving the Baxter equation. In the next section we will generalise this construction to completely generic integrable boundaries.

\section{Generic boundaries and the inhomogeneous Baxter equation}
\label{sec:generic-boundaries}

We now restore fully generic integrable boundary matrices. Most of the construction of Section~\ref{sec:fsov-correlators} carries over unchanged, so we only highlight what is different.

\subsection{The inhomogeneous Baxter equation}
\label{sec:TQ}

\paragraph{Convenient parameterisation.}Before spelling out the inhomogeneous Baxter equation explicitly it will be useful to introduce a convenient reparameterisation of the boundary parameters. Following \eg \cite{Niccoli:2022xdy}, we trade $(\zeta_\pm,\kappa_\pm)$ for new parameters
$\varphi_\pm,\psi_\pm$ defined by
\begin{equation}
 \sinh\varphi_\pm\cosh\psi_\pm=\frac{\sinh\zeta_\pm}{2\kappa_\pm},
 \qquad
 \cosh\varphi_\pm\sinh\psi_\pm=\frac{\cosh\zeta_\pm}{2\kappa_\pm}\,.
\label{eq:reparam}
\end{equation}
We then define the function $a_K(u)$ by
\begin{align}
 a_K(u)={}&
 \frac{\sinh(u-\tfrac\eta2+\varphi_-)}{\sinh\varphi_-}
 \frac{\sinh(u-\tfrac\eta2+\varphi_+)}{\sinh\varphi_+}
 \frac{\cosh(u-\tfrac\eta2+\psi_-)}{\cosh\psi_-}
 \frac{\cosh(u-\tfrac\eta2-\psi_+)}{\cosh\psi_+}\,,
\label{eq:aK}
\end{align}
which will be useful in what follows. Finally, we will use the function $\mathsf{A}(u)$ (we use the notation $\mathsf{A}(u)$ to distinguish it from the diagonal boundary function $A(u)$ of \eqref{eq:A-diagonal}) which takes the form
\begin{equation}
     \mathsf{A}(u) = (-1)^L\frac{\sinh(2u+\eta)}{\sinh(2u)}\,a_K(u)\,a(u)d(-u)\,.
\label{eq:A-generic}
\end{equation}
In the diagonal limit $\kappa_\pm\to0$, $\mathsf{A}(u)$ reduces to $A(u)$, since as  $\kappa_\pm\to0$ we see that $\varphi_\pm\to\zeta_\pm$
while $\psi_\pm\to\infty$, so that the two $\cosh$ factors in~\eqref{eq:aK}
cancel against each other.

\paragraph{Inhomogeneous Baxter equation.}
With the above definitions we can now write down the general Baxter equation.

As described in \secref{sec:openxxz}, for $\kappa_\pm\neq0$ the global $\alg{u}(1)$ symmetry is broken. The effect of this is that now each transfer matrix eigenvalue $\tau(u)$ is encoded in a Baxter equation with an inhomogeneous term $F(u)$~\cite{Nepomechie:2013ila,Cao:2013bca,Belliard:2018pvg,Belliard:2013aaa,Avan:2015ada,Kitanine:2014swa,Nepomechie:2020gwq}
\begin{equation}
 \mathsf{A}(u)Q(u-\eta)-\tau(u)Q(u)+\mathsf{A}(-u)Q(u+\eta)+F(u)=0\,,
\label{eq:TQ}
\end{equation}
where the inhomogeneous part $F(u)$ is explicitly given by
\begin{equation}
     F(u) = c_F\,a(u)a(-u)d(u)d(-u)\big[\cosh^2(2u)-\cosh^2\eta\big]\,,
\label{eq:F-generic}
\end{equation}
where we remind the reader of the definitions \eqref{eq:ad-bulk}
\begin{equation}
     a(u) = \sinh^L\big(u+\tfrac\eta2\big),\quad d(u) = \sinh^L\big(u-\tfrac\eta2\big)\,,
\end{equation}
and the constant $c_F$ is given by
\begin{equation}
     c_F=\frac{2\kappa_-\kappa_+}{\sinh\zeta_-\sinh\zeta_+}
     \Big[\cosh(\omega_--\omega_+)
     -\cosh\big(\varphi_-+\varphi_++\psi_--\psi_+-(L+1)\eta\big)\Big],
\label{eq:c-explicit}
\end{equation}
with $\varphi_\pm,\psi_\pm$ defined in \eqref{eq:reparam}.
The inhomogeneous term vanishes if and only if
$c_F=0$, which is not the case for generic boundaries.

Vanishing of this constant is one of the constraints imposed in \cite{Niccoli:2022xdy,Niccoli:2022mfr}.

\paragraph{Structure of Q-functions.}
Now we discuss the structure of the Q-functions which solve the Baxter equation.

In the diagonal case, the number $M$ which parameterised the global $\mathbb{S}^z$ eigenvalue $L/2-M$ (see \secref{sec:fsov-TQ-diagonal}) played a distinguished role and governed the large-$u$ asymptotics of the Q-function via
\begin{equation}
     Q(u) = \sinh^{2M}u + \dots
\end{equation}
where $\dots$ refers to lower order terms in $\sinh^2 u$. For generic boundaries this $\mathfrak{u}(1)$ symmetry is broken and the eigenstates $|\Psi\rangle$ are not organised into sectors of fixed $\mathbb{S}^z$. As a result, the asymptotics of the Q-function changes. Now, the Q-function for every state is a polynomial in $\sinh^2 u$ of degree $L$ (\ie the length of the chain)~\cite{Cao:2013bca,Belliard:2014rna,Kitanine:2014swa,Nepomechie:2013ila}\footnote{In our parameterisation, which is monic in $\sinh^2u$, the diagonal limit $\kappa_\pm\to0$ causes the Q-function to diverge. This is purely an artefact of normalisation: $F(u)\propto c_F$ vanishes at the same rate at which $Q(u)\sim1/c_F$ diverges, so rescaling $Q\to c_F\,Q$ and $F\to c_F\,F$ gives a Q-function that stays finite in the $\kappa_\pm\rightarrow 0$ limit, with a source that vanishes like $c_F^2$. This singular normalisation will drop out of all physical quantities of interest, \eg the boundary one-point functions.}
\begin{equation}
 Q(u)=\sinh^{2L}u+\sum_{j=0}^{L-1} c_j \sinh^{2j}u\,.
\label{eq:Qfn-generic}
\end{equation}
Remarkably, despite these new complications, the inhomogeneous Baxter equation continues to completely characterise the spectrum of the transfer matrix. In particular, for each eigenvalue $\tau(u)$ of the transfer matrix there is a unique such $Q(u)$. This is established in the presence of local impurities in~\cite{Kitanine:2014swa,Niccoli:2012pi,Faldella:2013qha} and we verified it here too for $L=1,2,3,4$.

\subsection{Functional SoV with an inhomogeneous Baxter equation}
\label{sec:fsov-generic-measure}

We now take the inhomogeneous Baxter equation and set up the Functional SoV formalism.

\paragraph{Self-adjointness and the inhomogeneous term.}
As in the diagonal boundary case of \secref{sec:fsov-measure}, we denote $Df(u):=f(u+\eta/2)$ and write \eqref{eq:TQ} as
\begin{equation}
 \mathcal O:=\mathsf{A}(u)D^{-2}+\mathsf{A}(-u)D^2-\tau(u),
 \qquad \mathcal O Q=-F.
\label{eq:fsov-baxterop}
\end{equation}
The inhomogeneous term $F$ does not enter the operator
$\mathcal O$, so the construction of the $L$ independent brackets
(\secref{sec:fsov-measure}) applies verbatim
with $A$ replaced by $\mathsf{A}$.
Indeed, the only properties of $A$ used there were its zero of order
$2L+1$ at $u=-\eta/2$ and $A(\eta/2)\neq0$. Both also hold for~$\mathsf{A}$
for generic boundary parameters, since the factors $a_K(u)$ and
$1/\sinh(2u)$ in~\eqref{eq:A-generic} are regular and non-zero at
$u=\pm\eta/2$.

We therefore use the same $L$ brackets
\begin{equation}
 \langle f\rangle_\alpha
 =\oint_{\gamma_+}{\rm d}u\,\mu_\alpha(u)\,f(u)
 +\oint_{\gamma_-}{\rm d}u\,\mu_\alpha(u)\,
 \frac{\mathsf{A}(u+\eta)}{\mathsf{A}(-u)}\,f(u)\,,
 \qquad \alpha=1,\dots,L\,,
\label{eq:fsov-bracket-alpha}
\end{equation}
with $\mu_\alpha$ given by \eqref{eq:mu-alpha-short},
and $\mathcal O$ is self-adjoint with respect to each of them.

We see that the inhomogeneous term $F(u)$ causes no difficulty in self-adjointness. Remarkably, the structure of $F(u)$ guarantees that it has no impact on the functional construction at all, thanks to the very useful property
\begin{equation}\label{eq:fsov-source-vanishing}
  \langle f F \rangle_\alpha=0,\quad \alpha=1,\dots,L\,,
\end{equation}
for any entire function $f$. Indeed, using $\cosh^2(2u)-\cosh^2\eta=\sinh(2u+\eta)\sinh(2u-\eta)$,
the source \eqref{eq:F-generic} can be written as
\begin{equation}
 F(u)=c_F\,\sinh^2(2u)\,
 \frac{\mathsf{A}(u)\,\mathsf{A}(-u)}{a_K(u)\,a_K(-u)}\,,
\label{eq:F-via-A}
\end{equation}
so it vanishes to order $2L+1$ at both $u=\pm\eta/2$.
These zeros cancel the poles of $\mu_\alpha$,
and hence \eqref{eq:fsov-source-vanishing} holds for any entire function $f$.

\paragraph{Functional scalar product.} We can now write down the functional scalar product.

Let $Q_A,Q_B$ solve \eqref{eq:TQ} for two transfer matrix eigenvalues $\tau_A,\tau_B$.
Now $\mathcal O_AQ_A=-F=\mathcal O_BQ_B$, and we find
\begin{equation}
 \big\langle Q_B(\mathcal O_A-\mathcal O_B)Q_A\big\rangle_\alpha
 =-\langle Q_BF\rangle_\alpha+\langle Q_AF\rangle_\alpha=0.
\label{eq:fsov-orthog-step}
\end{equation}
The first equality uses self-adjointness once, to write $\langle Q_B\mathcal O_BQ_A\rangle_\alpha
=\langle Q_A\mathcal O_BQ_B\rangle_\alpha$.
The second uses \eqref{eq:fsov-source-vanishing}.
Using the decomposition \eqref{eq:t-decomposition}, this reproduces the linear system of
Section~\ref{sec:fsov-orthogonality},
\begin{equation}
 \sum_{\beta=1}^L(H_\beta^A-H_\beta^B)M^{AB}_{\alpha\beta}=0,
 \qquad
 M^{AB}_{\alpha\beta}=\Big\langle Q_AQ_B\, p(u)
 \, s(u)^{\beta-1}\Big\rangle_\alpha\,.
\label{eq:orthogonality-tau-exp}
\end{equation}
As before this gives $\det M^{AB}\propto\delta_{AB}$. The remarkable conclusion is that the functional scalar product construction is completely general, with precisely the same scalar product working for both the diagonal boundary case and the generic boundary case. All of the differences between the two boundary configurations are completely contained in the Q-functions.

\subsection{Boundary one-point functions for generic boundaries}
\label{sec:fsov-generic-observables}

We are now ready to compute the boundary one-point functions for general integrable boundaries.

\paragraph{Derivatives of the conserved charges.}
First, as in \secref{sec:fsov-orthogonality},
we compute a general expression for the derivatives
of the integrals of motion with respect to a boundary parameter $\lambda$.
The computation from the diagonal boundary case goes through immediately:
differentiating $\mathcal O Q=-F$ and projecting onto $Q$ with the brackets,
the inhomogeneous term drops out by \eqref{eq:fsov-source-vanishing},
since $\partial_\lambda Q$ is again a polynomial and the boundary parameters enter $F$ only through the constant $c_F$, so that $\partial_\lambda F=(\partial_\lambda c_F/c_F)\,F$.
Hence $\langle Q(\partial_\lambda\mathcal O)Q\rangle_\alpha=0$ exactly as in~\eqref{eqn:diff-baxter}.
Writing
\begin{equation}
 \mathcal O=Y- p(u)\sum_{\beta=1}^LH_\beta\, s(u)^{\beta-1},
 \qquad
 Y=\mathsf{A}(u)D^{-2}+\mathsf{A}(-u)D^2-\mathcal C(u),
\label{eq:generic-Y}
\end{equation}
we obtain the same linear system as \eqref{eq:fsov-diag-derivative-system},
\begin{equation}
 \sum_{\beta=1}^L M_{\alpha\beta}\,\partial_\lambda H_\beta=y^{(\lambda)}_\alpha,
 \qquad
 M_{\alpha\beta}=\Big\langle Q^2 p(u)\, s(u)^{\beta-1}\Big\rangle_\alpha,
 \qquad
 y^{(\lambda)}_\alpha=\big\langle Q(\partial_\lambda\,Y)Q\big\rangle_\alpha,
\label{eq:formfactor-system}
\end{equation}
so that
\begin{equation}
 \partial_\lambda H_\beta=
 \frac{\det M^{(\beta,\lambda)}}{\det M},
\label{eq:formfactor}
\end{equation}
where $M^{(\beta,\lambda)}$ is $M$ with column $\beta$ replaced by $y^{(\lambda)}$.
By the Hellmann--Feynman theorem~\eqref{eq:hellmann-feynman}, this is the
diagonal matrix element of $\partial_\lambda\hat H_\beta$.

\paragraph{Boundary one-point functions: main result.}
\label{sec:fsov-generic-one-point-functions}
We can now assemble the main result of this paper.
The boundary spin operators were expressed in
\eqref{eq:boundary-magnetization} and \eqref{eq:boundary-magnetization-pm}
as boundary-parameter derivatives of the Hamiltonian.
For the left and right boundaries we introduce the derivative operators
\begin{equation}
 \mathcal D^z_\pm:=\partial_{\zeta_\pm}
 +\kappa_\pm\coth\zeta_\pm\,\partial_{\kappa_\pm}\,,
 \qquad
 \mathcal D^\pm_-:=\kappa_-\partial_{\kappa_-}\pm\partial_{\omega_-}\,,
 \qquad
 \mathcal D^\pm_+:=\kappa_+\partial_{\kappa_+}\pm\partial_{\omega_+}\,,
 \label{eq:generic-boundary-derivatives}
\end{equation}
so that
\begin{equation}
 \sigma^z_\pm=-\frac{\sinh^2\zeta_\pm}{\sinh\eta}\,\mathcal D^z_\pm\,\mathbb{H},
 \qquad
 \sigma^\pm_-=\frac{e^{\mp\omega_-}\sinh\zeta_-}{4\kappa_-\sinh\eta}\,
 \mathcal D^\pm_-\,\mathbb{H}\,,
 \qquad
 \sigma^\pm_+=\frac{e^{\mp\omega_+}\sinh\zeta_+}{4\kappa_+\sinh\eta}\,
 \mathcal D^\pm_+\,\mathbb{H}\,,
\end{equation}
where we remind the reader that the superscript refers to the direction of the spin operators while the subscript refers to the left or right boundary.
By the Hellmann--Feynman theorem \eqref{eq:hellmann-feynman}
the one-point functions are obtained by replacing $\mathbb{H}$ with its eigenvalue $E$.
By \eqref{eq:H-H1}, $E$ depends on the state only through $H_1$,
so for any of the derivatives $\mathcal D$ above
\begin{equation}
 \mathcal D E
 =\frac{\sinh\eta}{\mathcal C_1}
 \Big[\mathcal D\,\mathcal C'(\eta/2)
 +2\sinh(2\eta)\,\mathcal D H_1\Big],
 \qquad
 \mathcal D H_1=\frac{\det M^{(1,\mathcal D)}}{\det M}\,,
\label{eq:generic-DE}
\end{equation}
where the last equality is \eqref{eq:formfactor} with $\beta=1$,
and $M^{(1,\mathcal D)}$ denotes $M$ with its first column replaced by
\begin{equation}
 y^{\mathcal D}_\alpha:=\big\langle Q(\mathcal D\,Y)Q\big\rangle_\alpha\,.
\label{eq:generic-boundary-y}
\end{equation}
Putting everything together, the boundary one-point functions at the two boundaries read
\begin{equation}
\boxed{\begin{aligned}
 \big\langle\sigma^z_\pm\big\rangle_\Psi
 &= -\frac{\sinh^2\zeta_\pm}{\mathcal C_1}
 \left[\mathcal D^z_\pm\,\mathcal C'(\eta/2)
 +2\sinh(2\eta)\,\frac{\det M^{(1,\mathcal D^z_\pm)}}{\det M}\right],\\[6pt]
 \big\langle\sigma^{\pm}_-\big\rangle_\Psi
 &= \frac{e^{\mp\omega_-}\sinh\zeta_-}{4\kappa_-\,\mathcal C_1}
 \left[\mathcal D^\pm_-\,\mathcal C'(\eta/2)
 +2\sinh(2\eta)\,\frac{\det M^{(1,\mathcal D^\pm_-)}}{\det M}\right],\\[6pt]
 \big\langle\sigma^{\pm}_+\big\rangle_\Psi
 &= \frac{e^{\mp\omega_+}\sinh\zeta_+}{4\kappa_+\,\mathcal C_1}
 \left[\mathcal D^\pm_+\,\mathcal C'(\eta/2)
 +2\sinh(2\eta)\,\frac{\det M^{(1,\mathcal D^\pm_+)}}{\det M}\right].
\end{aligned}}
\label{eq:generic-boundary-spins}
\end{equation}
The first line has the same form as the diagonal result
\eqref{eq:fsov-boundarymag-compact}, now with the generic $\mathsf{A}$ in $Y$
and the Q-function solving the inhomogeneous Baxter equation.
The last two lines are new: the transverse one-point functions vanish identically
for diagonal boundaries but are non-trivial here.
All information about the inhomogeneous term of the Baxter equation enters only through the Q-function.

\paragraph{Transverse one-point functions.}
The derivative columns entering the transverse one-point functions simplify.
Split $Y$ into its shift part and its central part,
\begin{equation}
 Y=Y_{\mathsf A}-\mathcal C(u),
 \qquad
 Y_{\mathsf A}:=\mathsf{A}(u)D^{-2}+\mathsf{A}(-u)D^2\,.
\end{equation}
The function $\mathsf{A}$ does not depend on $\omega_\pm$,
so $\partial_{\omega_\pm}$ acts only on $\mathcal C(u)$.
Moreover, in \eqref{eq:central-full} the coefficients $\mathcal C_1$ and $\mathcal C_2$
depend on the boundary parameters only through $\zeta_\pm$,
while $\mathcal C_3$ also depends on $\kappa_\pm$ and $\omega_\pm$.
From \eqref{eq:leading-asymptotics} one finds
\begin{equation}
 \kappa_\pm\partial_{\kappa_\pm}\mathcal C_3=\mathcal C_3,
 \qquad
 \partial_{\omega_-}\mathcal C_3=\tanh(\omega_--\omega_+)\,\mathcal C_3\,,
 \qquad
 \partial_{\omega_+}\mathcal C_3=-\tanh(\omega_--\omega_+)\,\mathcal C_3\,,
\end{equation}
and therefore
\begin{align}
 \mathcal D^\pm_-\,\mathcal C(u)
 &=\frac{\mathcal C_3}{4}\Big[1\pm\tanh(\omega_--\omega_+)\Big]
 \big[\sinh^2(2u)-\sinh^2\eta\big]\sinh^{2L}u\,,\\
 \mathcal D^\pm_+\,\mathcal C(u)
 &=\frac{\mathcal C_3}{4}\Big[1\mp\tanh(\omega_--\omega_+)\Big]
 \big[\sinh^2(2u)-\sinh^2\eta\big]\sinh^{2L}u\,.
\end{align}
The derivative column \eqref{eq:generic-boundary-y} for $\mathcal D=\mathcal D^\pm_-$ and $\mathcal D=\mathcal D^\pm_+$ then becomes
\begin{align}
 y^{\mathcal D^\pm_-}_\alpha&=
 \big\langle Q\,(\kappa_-\partial_{\kappa_-}Y_{\mathsf A})\,Q\big\rangle_\alpha
 -\frac{\mathcal C_3}{4}\Big[1\pm\tanh(\omega_--\omega_+)\Big]
 \big\langle Q^2\,p(u)\sinh^{2L}u\big\rangle_\alpha\,,\\
 y^{\mathcal D^\pm_+}_\alpha&=
 \big\langle Q\,(\kappa_+\partial_{\kappa_+}Y_{\mathsf A})\,Q\big\rangle_\alpha
 -\frac{\mathcal C_3}{4}\Big[1\mp\tanh(\omega_--\omega_+)\Big]
 \big\langle Q^2\,p(u)\sinh^{2L}u\big\rangle_\alpha\,,
\end{align}
where we used $\sinh^2(2u)-\sinh^2\eta=p(u)$.
Hence, apart from the Q-function itself, the $\omega$-derivatives
enter the transverse one-point functions only through the explicit
prefactors in \eqref{eq:generic-boundary-spins} and through a single
additional moment
$\langle Q^2\,p(u)\sinh^{2L}u\rangle_\alpha$,
which is a bracket of exactly the same type as the entries of $M$.

As in the diagonal case, all brackets reduce to residues at $u=\pm\eta/2$,
so \eqref{eq:generic-boundary-spins} is a completely explicit expression
for the boundary one-point functions in terms of the Q-function of the state.
In the next section we test our expressions numerically.

\section{Finite-volume numerical checks}
\label{sec:generic-explicit-data}

In this section we test our main result \eqref{eq:generic-boundary-spins}
for the boundary one-point functions
against direct numerical diagonalisation.
Mathematica notebooks reproducing all of these checks for $L=1,2,3,4$ are provided as ancillary files with the arXiv submission of this paper. Below we summarise the results for $L=2,3$.

\subsection{Overview of numerics}

We carry out the numerical checks at two parameter points.

The first configuration, in \secref{sec:explicit-generic}, is a generic complex point.
All boundary parameters and the anisotropy are complex with no special relations among them,
the Hamiltonian is not Hermitian, and left and right eigenvectors genuinely differ.

The second, in \secref{sec:explicit-hermitian}, is a Hermitian configuration with purely imaginary $\eta=\ii$.
Writing $\Delta:=\cosh\eta$ for the coefficient of $\sigma^z_\alpha\sigma^z_{\alpha+1}$
in the Hamiltonian \eqref{eq:hamiltonian},
it lies in the gapless (critical) regime $|\Delta|<1$,
where open boundaries and boundary fields give rise to boundary critical behaviour
that has been studied extensively~\cite{Giamarchi:2003qpod,EggertAffleck:1992,Fujimoto:2004boundary,Sirker:2008impurities}.
Hermiticity implies reality and conjugation properties of the boundary one-point functions.
The fact that these are reproduced by the Functional SoV formula, even though Hermiticity is never used as input in the construction, serves as a useful cross-check.

\paragraph{Numerical procedure.}
All parameters are exact complex rationals,
and every numerical evaluation is carried out at a working precision of 60 digits.
For each parameter point and each $L$ we proceed as follows.

\emph{Direct side.}
We build the Hamiltonian \eqref{eq:hamiltonian} from Pauli matrices
and diagonalise it.
In all cases its spectrum is non-degenerate.
For each eigenstate we construct the left and right eigenvectors
$\langle\Psi|$ and $|\Psi\rangle$.
The left eigenvector is an eigenvector of the transpose $\mathbb H^{\mathsf T}$,
with no complex conjugation,
in line with \eqref{eqn:lefteigenvec}.
We normalise the right eigenvectors to unit Euclidean norm,
$\sum_i|\Psi^i_A|^2=1$ in the basis of $\sigma^z$ eigenstates,
and fix the scale of each left eigenvector by $\langle\Psi_A|\Psi_A\rangle=1$.
Together with orthogonality of distinct eigenstates,
this gives $\langle\Psi_A|\Psi_B\rangle=\delta_{AB}$.
At the Hermitian point the left eigenvector is then
the complex conjugate of the right one,
$\langle\Psi_A|=\overline{|\Psi_A\rangle}^{\mathsf T}$.
None of the quantities compared below depends on these choices,
since the one-point functions are normalised by $\langle\Psi|\Psi\rangle$.
We then compute the boundary one-point functions
$\langle\sigma^a_\pm\rangle_\Psi$, $a\in\{z,\pm\}$, directly.

\emph{FSoV side.}
For each state we compute the transfer matrix eigenvalue $\tau(u)$.
Evaluating $\tau(u)$ at $L$ points and using \eqref{eq:t-decomposition}
gives the integrals of motion $H_1,\dots,H_L$.
Next, we feed $\tau(u)$ into the Baxter equation \eqref{eq:TQ},
with the explicit source constant \eqref{eq:c-explicit},
and write the Q-function as in \eqref{eq:Qfn-generic}.
Both sides of \eqref{eq:TQ} are even in $u$.
Expanding around $u=0$, the coefficients of $u^0,u^2,\dots,u^{2L-2}$
give $L$ linear equations for the $L$ coefficients $c_0,\dots,c_{L-1}$ of the Q-function.
We then evaluate the brackets $\langle-\rangle_\alpha$ \eqref{eq:fsov-bracket-alpha}
as residues at $u=\pm\eta/2$.
Since the measures \eqref{eq:mu-alpha-short} are elementary functions,
this step is exact and applies for any $\eta$.
Finally, we insert the brackets into \eqref{eq:generic-boundary-spins}.
The boundary derivatives \eqref{eq:generic-boundary-derivatives}
of $\mathsf A(u)$ and $\mathcal C(u)$ are taken symbolically,
so no finite differences enter.
Beyond $\tau(u)$, no information from the diagonalisation is used on the FSoV side.

\paragraph{Consistency checks.}

The Q-functions are fixed by the lowest $L$ orders of the expansion
around $u=0$. The higher orders (up to $u^{20}$) therefore
independently test both the solution and the explicit value of $c_F$.
They are satisfied to more than $40$ significant digits in all cases.

We also test the orthogonality of distinct eigenstates using FSoV quantities alone.
The functional construction predicts $\det M^{AB}\propto\delta_{AB}$,
see \eqref{eq:orthogonality-tau-exp},
but does not fix the proportionality constant.
The combination
\begin{equation}
 G_{AB}:=\frac{|\det M^{AB}|}{|\det M^{AA}\,\det M^{BB}|^{1/2}}
\label{eq:explicit-overlap}
\end{equation}
is independent of the normalisation of the brackets,
and is computed from the Q-functions and the brackets alone,
without any eigenvector.
By construction $G_{AA}=1$, and orthogonality predicts $G_{AB}=0$ for $A\neq B$.
The values are listed in Tables~\ref{tab:generic-overlap-L2} and~\ref{tab:generic-overlap-L3} for the generic point,
and in Tables~\ref{tab:hermitian-overlap-L2} and~\ref{tab:hermitian-overlap-L3} for the Hermitian point.

We also confirm the assumption of \secref{sec:fsov-orthogonality} that
$\det M^{AA}\neq0$.
Since $\det M^{AA}$ depends on the normalisation of the Q-function and of the brackets,
we consider the ratio
\begin{equation}
 h_A:=\frac{|\det M^{AA}|}{\prod_{\alpha=1}^L\|M^{AA}_{\alpha\cdot}\|}\,,
 \qquad
 \|M_{\alpha\cdot}\|:=\Big(\sum_{\beta=1}^L|M_{\alpha\beta}|^2\Big)^{1/2},
\label{eq:hadamard-ratio}
\end{equation}
of the determinant to the product of the row norms.
It is invariant under rescalings of $Q_A$ and of the individual brackets,
satisfies $0\leq h_A\leq1$,
and vanishes if and only if $M^{AA}$ is singular.
For all states at both parameter points, $h_A$ lies between $6\times10^{-9}$ and $2\times10^{-2}$,
while $\det M^{AA}$ itself is determined to at least 47 significant digits,
see \tabref{tab:detM}.

The three assumptions of our construction,
namely a non-degenerate spectrum, a unique Q-function for every
transfer matrix eigenvalue, and $\det M^{AA}\neq0$,
are thus confirmed for $L=2,3$ at both parameter points.
We have checked them in the same way also for $L=1$ and $L=4$,
and for diagonal boundaries.

\begin{table}
 \centering
 \small
 \begin{tabular}{lcccc}
  \toprule
  & \multicolumn{2}{c}{generic point} & \multicolumn{2}{c}{Hermitian point} \\ \cmidrule(lr){2-3}\cmidrule(lr){4-5}
  & $L=2$ & $L=3$ & $L=2$ & $L=3$ \\
  \midrule
  $\min_A h_A$ & $2.2\times10^{-3}$ & $6.3\times10^{-9}$ & $5.0\times10^{-3}$ & $5.1\times10^{-7}$ \\
  $\max_A h_A$ & $9.0\times10^{-3}$ & $1.1\times10^{-6}$ & $2.0\times10^{-2}$ & $1.0\times10^{-5}$ \\
  significant digits of $\det M^{AA}$ & $51$ & $47$ & $52$ & $48$ \\
  \bottomrule
 \end{tabular}
 \caption{Check of the assumption $\det M^{AA}\neq0$.
 The ratio $h_A$ of \protect\eqref{eq:hadamard-ratio}, minimised and maximised over all states,
 and the minimal number (over all states) of correct significant digits of $\det M^{AA}$
 at a working precision of 60 digits.}
\label{tab:detM}
\end{table}

\paragraph{Precision and accuracy.}
All inputs are exact, and numerical evaluation uses 60 significant digits.
Mathematica tracks how many digits of every intermediate result are correct,
and some digits are lost along the way,
mostly on the FSoV side in solving for the Q-functions
and in the high-order Taylor coefficients entering the residues. Repeating the computation at the generic point
with 100 instead of 60 digits
improves all bounds on the one-point functions by 40 orders of magnitude,
to $10^{-89}$ for $L=2$ and $10^{-85}$ for $L=3$.
In the tables below, energies are rounded to 8 decimals,
and one-point functions to 4 decimals at the generic point and to 6 decimals at the Hermitian point,
so that the tables fit on the page.
The full-precision values are available in the ancillary notebooks.

\subsection{Generic configuration}
\label{sec:explicit-generic}

\paragraph{Parameters.}
We take $\eta=\tfrac12+\tfrac{\ii}{3}$ together with
\begin{equation}
 (\zeta_-,\kappa_-,\omega_-)=\big(\tfrac35+\tfrac{\ii}{4},\tfrac25-\tfrac{\ii}{7},\tfrac16+\tfrac{\ii}{5}\big),
 \qquad
 (\zeta_+,\kappa_+,\omega_+)=\big(\tfrac27-\tfrac{\ii}{3},\tfrac13+\tfrac{\ii}{8},-\tfrac14+\tfrac{\ii}{9}\big),
\label{eq:generic-explicit-parameters}
\end{equation}
for both chain lengths.
The Hamiltonian is not Hermitian and its spectrum is complex,
so left and right eigenvectors genuinely differ.
The parameters \eqref{eq:reparam} entering the Baxter equation are
\begin{equation}
 \begin{aligned}
 \varphi_-&=0.4365550930+0.2477058892\,\ii,
 &\psi_-&=1.0428054404+0.2981795933\,\ii,\\
 \varphi_+&=0.2091314511-0.3034861012\,\ii,
 &\psi_+&=1.1290612703-0.3219520380\,\ii,
 \end{aligned}
\end{equation}
and the source constant \eqref{eq:c-explicit} evaluates to
\begin{equation}
 c_F\big|_{L=2}=-0.03901457-0.49598066\,\ii,
 \qquad
 c_F\big|_{L=3}=0.14401751-1.45639160\,\ii.
\end{equation}
Since $c_F\neq0$, the source term $F(u)$ does not vanish,
and this configuration lies away from the constrained surfaces
of~\cite{Niccoli:2022mfr} where the Baxter equation becomes homogeneous.

\paragraph{States.}
We label the $2^L$ eigenstates $|\Psi_A\rangle$ by $A=1,\dots,2^L$,
in order of increasing real part of the energy.
Their energies are listed in \tabref{tab:generic-energies}.

\begin{table}
 \centering
 \small
 \begin{tabular}{crr}
  \toprule
  $A$ & $E_A$, $L=2$ & $E_A$, $L=3$ \\
  \midrule
  $1$ & $-3.18111626+0.74644511\,\ii$ & $-4.72264244-1.59275694\,\ii$ \\
  $2$ & $-0.07645228-1.87510888\,\ii$ & $-3.38008023+1.34370775\,\ii$ \\
  $3$ & $ 0.96581905-0.88485981\,\ii$ & $-0.78185313+0.83310586\,\ii$ \\
  $4$ & $ 2.29174950+2.01352358\,\ii$ & $ 0.54164705+0.75676314\,\ii$ \\
  $5$ &                             &   $ 0.96098216-1.64914863\,\ii$ \\
  $6$ &                             &   $ 2.01245079-0.33106759\,\ii$ \\
  $7$ &                             &   $ 2.04708798-1.48891118\,\ii$ \\
  $8$ &                             &   $ 3.32240780+2.12830759\,\ii$ \\
  \bottomrule
 \end{tabular}
 \caption{Energies of the $2^L$ eigenstates at the generic point~\protect\eqref{eq:generic-explicit-parameters}.
 The states are labelled by $A=1,\dots,2^L$ in order of increasing $\operatorname{Re}E_A$.
 This labelling is used in all tables of \secref{sec:explicit-generic}.}
\label{tab:generic-energies}
\end{table}

\paragraph{Results.}
For every eigenstate and all six observables
$\langle\sigma^z_\pm\rangle_\Psi$, $\langle\sigma^+_\pm\rangle_\Psi$, $\langle\sigma^-_\pm\rangle_\Psi$,
the FSoV formula \eqref{eq:generic-boundary-spins}
and the direct computation agree,
to better than $10^{-49}$ for $L=2$ and $10^{-45}$ for $L=3$.
The values are listed in
Tables~\ref{tab:generic-onepoint-L2} and~\ref{tab:generic-onepoint-L3}.

\paragraph{Orthogonality.}
The normalised overlaps \eqref{eq:explicit-overlap},
computed from FSoV alone,
are listed in Tables~\ref{tab:generic-overlap-L2} and~\ref{tab:generic-overlap-L3}.
All off-diagonal values are below $10^{-54}$,
that is, $\det M^{AB}$ vanishes at the working precision of 60 digits.

\begin{table}
 \centering
 \small
 \setlength{\tabcolsep}{3pt}
 \begin{tabular}{ccccc}
  \toprule
  $A\backslash B$ & $1$ & $2$ & $3$ & $4$ \\
  \midrule
$1$ & $1$ & $8.4\times10^{-58}$ & $4.3\times10^{-58}$ & $9.1\times10^{-58}$ \\
$2$ &  & $1$ & $1.7\times10^{-57}$ & $1.0\times10^{-57}$ \\
$3$ &  &  & $1$ & $7.2\times10^{-58}$ \\
$4$ &  &  &  & $1$ \\
  \bottomrule
 \end{tabular}
 \caption{Normalised overlaps $G_{AB}$, eq.~\protect\eqref{eq:explicit-overlap}, computed from FSoV alone,
 for $L=2$ at the generic point~\protect\eqref{eq:generic-explicit-parameters}.
 The matrix is symmetric, and $G_{AA}=1$ by construction.}
\label{tab:generic-overlap-L2}
\end{table}

\begin{table}
 \centering
 \scriptsize
 \setlength{\tabcolsep}{3pt}
 \begin{tabular}{ccccccccc}
  \toprule
  $A\backslash B$ & $1$ & $2$ & $3$ & $4$ & $5$ & $6$ & $7$ & $8$ \\
  \midrule
$1$ & $1$ & $4.2\times10^{-56}$ & $1.3\times10^{-55}$ & $4.6\times10^{-56}$ & $7.9\times10^{-56}$ & $7.4\times10^{-56}$ & $7.1\times10^{-56}$ & $7.5\times10^{-57}$ \\
$2$ &  & $1$ & $1.4\times10^{-55}$ & $1.0\times10^{-55}$ & $6.9\times10^{-56}$ & $7.4\times10^{-56}$ & $4.4\times10^{-56}$ & $4.3\times10^{-56}$ \\
$3$ &  &  & $1$ & $1.4\times10^{-55}$ & $3.2\times10^{-55}$ & $1.9\times10^{-55}$ & $1.6\times10^{-55}$ & $1.3\times10^{-55}$ \\
$4$ &  &  &  & $1$ & $1.4\times10^{-55}$ & $3.3\times10^{-55}$ & $2.7\times10^{-55}$ & $9.7\times10^{-56}$ \\
$5$ &  &  &  &  & $1$ & $4.7\times10^{-55}$ & $4.5\times10^{-55}$ & $1.4\times10^{-55}$ \\
$6$ &  &  &  &  &  & $1$ & $6.8\times10^{-55}$ & $1.5\times10^{-55}$ \\
$7$ &  &  &  &  &  &  & $1$ & $1.9\times10^{-55}$ \\
$8$ &  &  &  &  &  &  &  & $1$ \\
  \bottomrule
 \end{tabular}
 \caption{Normalised overlaps $G_{AB}$, eq.~\protect\eqref{eq:explicit-overlap}, computed from FSoV alone,
 for $L=3$ at the generic point~\protect\eqref{eq:generic-explicit-parameters}.
 The matrix is symmetric, and $G_{AA}=1$ by construction.}
\label{tab:generic-overlap-L3}
\end{table}

\begin{table}
 \centering
 \scriptsize
 \setlength{\tabcolsep}{2pt}
 \begin{tabular}{crrrrrr}
  \toprule
  $A$ & $\langle\sigma^z_-\rangle$ & $\langle\sigma^+_-\rangle$ & $\langle\sigma^-_-\rangle$
  & $\langle\sigma^z_+\rangle$ & $\langle\sigma^+_+\rangle$ & $\langle\sigma^-_+\rangle$ \\
  \midrule
$1$ & $-0.5661+0.2937\,\ii$ & $-0.2576+0.1672\,\ii$ & $-0.2482+0.0557\,\ii$ & $0.4639-0.3253\,\ii$ & $0.3238-0.2144\,\ii$ & $0.2747-0.0144\,\ii$ \\
$2$ & $-0.9790+0.0090\,\ii$ & $-0.0078-0.0305\,\ii$ & $-0.2362+0.2297\,\ii$ & $-0.7602-0.0747\,\ii$ & $-0.2865+0.0887\,\ii$ & $-0.3626-0.0167\,\ii$ \\
$3$ & $0.6790-0.3378\,\ii$ & $0.0253-0.0925\,\ii$ & $0.2212-0.2703\,\ii$ & $-0.6230+0.4608\,\ii$ & $-0.3042+0.0244\,\ii$ & $-0.0759-0.0112\,\ii$ \\
$4$ & $0.8662+0.0351\,\ii$ & $0.2401-0.0441\,\ii$ & $0.2633-0.0151\,\ii$ & $0.9193-0.0608\,\ii$ & $0.2668+0.1013\,\ii$ & $0.1638+0.0424\,\ii$ \\
  \bottomrule
 \end{tabular}
 \caption{$L=2$ boundary one-point functions
 for the generic parameters~\protect\eqref{eq:generic-explicit-parameters}.
 States are labelled as in \tabref{tab:generic-energies}.
 FSoV and direct diagonalisation agree to better than $10^{-49}$.}
\label{tab:generic-onepoint-L2}
\end{table}

\begin{table}
 \centering
 \scriptsize
 \setlength{\tabcolsep}{2pt}
 \begin{tabular}{crrrrrr}
  \toprule
  $A$ & $\langle\sigma^z_-\rangle$ & $\langle\sigma^+_-\rangle$ & $\langle\sigma^-_-\rangle$
  & $\langle\sigma^z_+\rangle$ & $\langle\sigma^+_+\rangle$ & $\langle\sigma^-_+\rangle$ \\
  \midrule
$1$ & $-0.6840-0.0061\,\ii$ & $-0.2124+0.0369\,\ii$ & $-0.2170+0.0194\,\ii$ & $-0.8284-0.2203\,\ii$ & $-0.2145-0.1948\,\ii$ & $-0.1179-0.0882\,\ii$ \\
$2$ & $0.1777+0.0172\,\ii$ & $0.0104+0.1495\,\ii$ & $-0.0933+0.0244\,\ii$ & $0.7981-0.1158\,\ii$ & $0.2480-0.0184\,\ii$ & $0.2164-0.0348\,\ii$ \\
$3$ & $-0.7490+0.0313\,\ii$ & $-0.2311-0.0144\,\ii$ & $-0.2004-0.0028\,\ii$ & $0.6275+0.0826\,\ii$ & $0.4107+0.0913\,\ii$ & $0.1941+0.1844\,\ii$ \\
$4$ & $-0.2864+0.0676\,\ii$ & $-0.0796+0.0572\,\ii$ & $-0.0494+0.0021\,\ii$ & $0.7416-0.8221\,\ii$ & $0.4743-0.3929\,\ii$ & $0.3675-0.1406\,\ii$ \\
$5$ & $-0.8733-0.1377\,\ii$ & $-0.1479+0.1120\,\ii$ & $-0.4696+0.0511\,\ii$ & $-0.6705+0.0939\,\ii$ & $-0.4250+0.0600\,\ii$ & $-0.3123-0.1179\,\ii$ \\
$6$ & $1.0564-0.1978\,\ii$ & $0.2723-0.0049\,\ii$ & $0.2504-0.0372\,\ii$ & $-0.6944+0.9905\,\ii$ & $-0.5137+0.3885\,\ii$ & $-0.3067+0.0699\,\ii$ \\
$7$ & $0.4800+0.1447\,\ii$ & $0.1672-0.2417\,\ii$ & $0.4971-0.0164\,\ii$ & $-0.9085+0.0698\,\ii$ & $-0.2550-0.0485\,\ii$ & $-0.1911+0.0558\,\ii$ \\
$8$ & $0.8786+0.0809\,\ii$ & $0.2211-0.0945\,\ii$ & $0.2821-0.0405\,\ii$ & $0.9346-0.0786\,\ii$ & $0.2751+0.1148\,\ii$ & $0.1499+0.0714\,\ii$ \\
  \bottomrule
 \end{tabular}
 \caption{$L=3$ boundary one-point functions
 for the generic parameters~\protect\eqref{eq:generic-explicit-parameters}.
 States are labelled as in \tabref{tab:generic-energies}.
 FSoV and direct diagonalisation agree to better than $10^{-45}$.}
\label{tab:generic-onepoint-L3}
\end{table}

\subsection{Hermitian configuration}
\label{sec:explicit-hermitian}

\paragraph{Parameters.}
We take $\eta=\ii$ together with
\begin{equation}
 (\zeta_-,\kappa_-,\omega_-)=\big(\tfrac{\ii}{2},\tfrac13,\tfrac{\ii}{5}\big),
 \qquad
 (\zeta_+,\kappa_+,\omega_+)=\big(\tfrac{\ii}{3},\tfrac14,-\tfrac{\ii}{7}\big),
\label{eq:hermitian-parameters}
\end{equation}
for both chain lengths.
The anisotropy parameter $\eta$ is purely imaginary, $\Delta=\cosh\eta=\cos1$,
so the chain is in the gapless regime.
For $\eta,\zeta_\pm,\omega_\pm\in\ii\mathbb R$ and $\kappa_\pm\in\mathbb R$
the Hamiltonian \eqref{eq:hamiltonian} is Hermitian.
The parameters \eqref{eq:reparam} entering the Baxter equation are
\begin{equation}
 \begin{aligned}
 \varphi_-&=0.4220375343\,\ii,
 &\psi_-&=1.1627162696,\\
 \varphi_+&=0.2997031129\,\ii,
 &\psi_+&=1.4337933344,
 \end{aligned}
\end{equation}
and the source constant \eqref{eq:c-explicit} evaluates to
\begin{equation}
 c_F\big|_{L=2}=-1.71668392+0.22158537\,\ii,
 \qquad
 c_F\big|_{L=3}=-2.09212770-0.03972189\,\ii.
\end{equation}
Again $F(u)\neq0$.

\paragraph{States.}
The energies are real here,
and we label the $2^L$ eigenstates $|\Psi_A\rangle$ by $A=1,\dots,2^L$ in order of increasing energy.
Their energies are listed in \tabref{tab:hermitian-energies}.

\begin{table}
 \centering
 \small
 \begin{tabular}{crr}
  \toprule
  $A$ & $E_A$, $L=2$ & $E_A$, $L=3$ \\
  \midrule
  $1$ & $-4.09627538$ & $-7.05224579$ \\
  $2$ & $-2.63872516$ & $-3.38734614$ \\
  $3$ & $1.37378135$ & $-1.79846988$ \\
  $4$ & $5.36121919$ & $-0.36324803$ \\
  $5$ &  & $0.24756738$ \\
  $6$ &  & $1.48925643$ \\
  $7$ &  & $4.65521016$ \\
  $8$ &  & $6.20927586$ \\
  \bottomrule
 \end{tabular}
 \caption{Energies of the $2^L$ eigenstates at the Hermitian point~\protect\eqref{eq:hermitian-parameters}.
 The energies are real, and the states are labelled by $A=1,\dots,2^L$ in order of increasing $E_A$.
 This labelling is used in all tables of \secref{sec:explicit-hermitian}.}
\label{tab:hermitian-energies}
\end{table}

\paragraph{Results.}
The FSoV formula \eqref{eq:generic-boundary-spins}
and the direct computation agree for every eigenstate and all six observables,
to better than $10^{-51}$ for $L=2$ and $10^{-47}$ for $L=3$.
The one-point functions are listed in
Tables~\ref{tab:hermitian-onepoint-L2} and~\ref{tab:hermitian-onepoint-L3}.

\paragraph{Orthogonality.}
The normalised overlaps \eqref{eq:explicit-overlap},
computed from FSoV alone,
are listed in Tables~\ref{tab:hermitian-overlap-L2} and~\ref{tab:hermitian-overlap-L3}.
Again all off-diagonal values are below $10^{-54}$,
so that $\det M^{AB}$ vanishes at the working precision for $A\neq B$.

\begin{table}
 \centering
 \small
 \setlength{\tabcolsep}{3pt}
 \begin{tabular}{ccccc}
  \toprule
  $A\backslash B$ & $1$ & $2$ & $3$ & $4$ \\
  \midrule
$1$ & $1$ & $4.7\times10^{-58}$ & $1.8\times10^{-58}$ & $5.5\times10^{-58}$ \\
$2$ &  & $1$ & $4.4\times10^{-58}$ & $1.4\times10^{-57}$ \\
$3$ &  &  & $1$ & $4.8\times10^{-57}$ \\
$4$ &  &  &  & $1$ \\
  \bottomrule
 \end{tabular}
 \caption{Normalised overlaps $G_{AB}$, eq.~\protect\eqref{eq:explicit-overlap}, computed from FSoV alone,
 for $L=2$ at the Hermitian point~\protect\eqref{eq:hermitian-parameters}.
 The matrix is symmetric, and $G_{AA}=1$ by construction.}
\label{tab:hermitian-overlap-L2}
\end{table}

\begin{table}
 \centering
 \scriptsize
 \setlength{\tabcolsep}{3pt}
 \begin{tabular}{ccccccccc}
  \toprule
  $A\backslash B$ & $1$ & $2$ & $3$ & $4$ & $5$ & $6$ & $7$ & $8$ \\
  \midrule
$1$ & $1$ & $8.6\times10^{-59}$ & $2.0\times10^{-58}$ & $3.9\times10^{-58}$ & $3.4\times10^{-58}$ & $1.0\times10^{-56}$ & $4.1\times10^{-57}$ & $7.2\times10^{-58}$ \\
$2$ &  & $1$ & $7.1\times10^{-58}$ & $4.1\times10^{-58}$ & $8.1\times10^{-58}$ & $3.7\times10^{-56}$ & $1.8\times10^{-56}$ & $4.1\times10^{-57}$ \\
$3$ &  &  & $1$ & $4.0\times10^{-57}$ & $3.3\times10^{-58}$ & $4.1\times10^{-56}$ & $1.2\times10^{-55}$ & $3.4\times10^{-56}$ \\
$4$ &  &  &  & $1$ & $2.6\times10^{-57}$ & $8.1\times10^{-57}$ & $1.8\times10^{-55}$ & $7.0\times10^{-56}$ \\
$5$ &  &  &  &  & $1$ & $5.2\times10^{-56}$ & $2.8\times10^{-55}$ & $8.5\times10^{-56}$ \\
$6$ &  &  &  &  &  & $1$ & $3.2\times10^{-55}$ & $1.1\times10^{-55}$ \\
$7$ &  &  &  &  &  &  & $1$ & $5.5\times10^{-55}$ \\
$8$ &  &  &  &  &  &  &  & $1$ \\
  \bottomrule
 \end{tabular}
 \caption{Normalised overlaps $G_{AB}$, eq.~\protect\eqref{eq:explicit-overlap}, computed from FSoV alone,
 for $L=3$ at the Hermitian point~\protect\eqref{eq:hermitian-parameters}.
 The matrix is symmetric, and $G_{AA}=1$ by construction.}
\label{tab:hermitian-overlap-L3}
\end{table}

\paragraph{Consequences of Hermiticity.}
At a Hermitian point
the energies are real, $\langle\sigma^z_\pm\rangle_\Psi$ is real,
and $\langle\sigma^-_\pm\rangle_\Psi=\overline{\langle\sigma^+_\pm\rangle_\Psi}$,
since $\sigma^-=(\sigma^+)^\dagger$ and $\langle\Psi|$ is the conjugate of $|\Psi\rangle$.
The FSoV construction does not know about Hermiticity:
the coefficients of the Q-functions and the individual brackets are complex.
Nevertheless, $\langle\sigma^z_\pm\rangle_\Psi$
computed from \eqref{eq:generic-boundary-spins} is real to working precision,
and $\langle\sigma^+_\pm\rangle_\Psi$ and $\langle\sigma^-_\pm\rangle_\Psi$,
computed independently from \eqref{eq:generic-boundary-spins},
are complex conjugates of each other to working precision.
Both properties are visible by inspection in
Tables~\ref{tab:hermitian-onepoint-L2} and~\ref{tab:hermitian-onepoint-L3}.

\begin{table}
 \centering
 \scriptsize
 \setlength{\tabcolsep}{2.5pt}
 \begin{tabular}{crrrrrr}
  \toprule
  $A$ & $\langle\sigma^z_-\rangle$ & $\langle\sigma^+_-\rangle$ & $\langle\sigma^-_-\rangle$
  & $\langle\sigma^z_+\rangle$ & $\langle\sigma^+_+\rangle$ & $\langle\sigma^-_+\rangle$ \\
  \midrule
$1$ & $-0.714946$ & $-0.304577+0.170229\,\ii$ & $-0.304577-0.170229\,\ii$ & $-0.925690$ & $-0.145660-0.118700\,\ii$ & $-0.145660+0.118700\,\ii$ \\
$2$ & $0.186761$ & $0.168245-0.153985\,\ii$ & $0.168245+0.153985\,\ii$ & $-0.340295$ & $-0.156446+0.085509\,\ii$ & $-0.156446-0.085509\,\ii$ \\
$3$ & $-0.224126$ & $-0.190345+0.021782\,\ii$ & $-0.190345-0.021782\,\ii$ & $0.440431$ & $0.020956+0.018153\,\ii$ & $0.020956-0.018153\,\ii$ \\
$4$ & $0.752311$ & $0.326677-0.038027\,\ii$ & $0.326677+0.038027\,\ii$ & $0.825555$ & $0.281150+0.015037\,\ii$ & $0.281150-0.015037\,\ii$ \\
  \bottomrule
 \end{tabular}
 \caption{$L=2$ boundary one-point functions computed from FSoV
 for the parameters~\protect\eqref{eq:hermitian-parameters}.
 States are labelled as in \tabref{tab:hermitian-energies}.
 FSoV and direct diagonalisation agree to better than $10^{-51}$.}
\label{tab:hermitian-onepoint-L2}
\end{table}

\begin{table}
 \centering
 \scriptsize
 \setlength{\tabcolsep}{2.5pt}
 \begin{tabular}{crrrrrr}
  \toprule
  $A$ & $\langle\sigma^z_-\rangle$ & $\langle\sigma^+_-\rangle$ & $\langle\sigma^-_-\rangle$
  & $\langle\sigma^z_+\rangle$ & $\langle\sigma^+_+\rangle$ & $\langle\sigma^-_+\rangle$ \\
  \midrule
$1$ & $-0.637067$ & $-0.279204+0.035658\,\ii$ & $-0.279204-0.035658\,\ii$ & $-0.753910$ & $-0.250420-0.017153\,\ii$ & $-0.250420+0.017153\,\ii$ \\
$2$ & $-0.861139$ & $-0.229570+0.073347\,\ii$ & $-0.229570-0.073347\,\ii$ & $-0.934258$ & $-0.162414-0.047517\,\ii$ & $-0.162414+0.047517\,\ii$ \\
$3$ & $0.660796$ & $0.228486+0.010550\,\ii$ & $0.228486-0.010550\,\ii$ & $-0.453398$ & $-0.051913-0.058703\,\ii$ & $-0.051913+0.058703\,\ii$ \\
$4$ & $0.554699$ & $0.210678-0.106734\,\ii$ & $0.210678+0.106734\,\ii$ & $-0.685261$ & $-0.204072+0.028336\,\ii$ & $-0.204072-0.028336\,\ii$ \\
$5$ & $-0.384511$ & $-0.147278+0.070994\,\ii$ & $-0.147278-0.070994\,\ii$ & $0.533999$ & $0.152061-0.015195\,\ii$ & $0.152061+0.015195\,\ii$ \\
$6$ & $-0.725709$ & $-0.244696+0.001009\,\ii$ & $-0.244696-0.001009\,\ii$ & $0.664906$ & $0.101767+0.058421\,\ii$ & $0.101767-0.058421\,\ii$ \\
$7$ & $0.753206$ & $0.088335-0.037238\,\ii$ & $0.088335+0.037238\,\ii$ & $0.858812$ & $0.102003+0.032090\,\ii$ & $0.102003-0.032090\,\ii$ \\
$8$ & $0.639725$ & $0.373250-0.047585\,\ii$ & $0.373250+0.047585\,\ii$ & $0.769109$ & $0.312989+0.019722\,\ii$ & $0.312989-0.019722\,\ii$ \\
  \bottomrule
 \end{tabular}
 \caption{$L=3$ boundary one-point functions computed from FSoV
 for the parameters~\protect\eqref{eq:hermitian-parameters}.
 States are labelled as in \tabref{tab:hermitian-energies}.
 FSoV and direct diagonalisation agree to better than $10^{-47}$.}
\label{tab:hermitian-onepoint-L3}
\end{table}

\section{Conclusions and outlook}
\label{sec:conclusions}

In this paper we have computed boundary one-point functions
in the open XXZ spin chain with the most general integrable
boundary conditions, at finite length $L$ and for arbitrary
transfer matrix eigenstates. Our main result
\eqref{eq:generic-boundary-spins} expresses the longitudinal
boundary magnetisations $\langle\sigma^z_\pm\rangle_\Psi$ and the
transverse one-point functions $\langle\sigma^\pm_\pm\rangle_\Psi$
as ratios of $L\times L$ determinants built from the Baxter
Q-function of the state. No constraint relating the two boundaries
is imposed, and no reference state is needed. The formulae therefore
hold in the regime where the $\alg{u}(1)$ symmetry is broken and the
Baxter equation is genuinely inhomogeneous, which so far has not been
reached by methods for local observables.

The construction rests on three simple ingredients. First, the
boundary spin operators are derivatives of the Hamiltonian with
respect to the boundary parameters, so by the Hellmann--Feynman
theorem their expectation values follow from derivatives of the
energy. Second, Functional Separation of Variables gives these
derivatives through Cramer's rule, without constructing eigenvectors
and without differentiating the Q-function. Third, the source term
of the inhomogeneous Baxter equation drops out of every functional
bracket, $\langle QF\rangle_\alpha=0$, because $F(u)$ vanishes
to order $2L+1$ at $u=\pm\eta/2$ and thereby cancels the poles
of the measure. This is
the functional counterpart of the vanishing of $F(u)$ at the
separated-variable points observed in the operatorial
approach~\cite{Kitanine:2014swa}. As a consequence, the measure,
the scalar product and the derivative formulae carry over from
diagonal to generic boundaries essentially unchanged. The
boundary parameters enter only through the Q-function, the
coefficient function $\mathsf{A}(u)$, and the state-independent
part $\mathcal C(u)$ of the transfer matrix eigenvalue.

We tested our main result \eqref{eq:generic-boundary-spins} against direct
diagonalisation for $L=2$ and $L=3$, both at a generic complex
point, where the Hamiltonian is not Hermitian, and at a Hermitian
point in the critical regime, with $\eta$ purely imaginary so that
$|\Delta|=|\cosh\eta|<1$. In all cases we found
agreement to the full working precision. At the Hermitian point,
the reality and conjugation properties of the one-point functions
are reproduced even though Hermiticity is never used as input.

Two statements in our construction are assumed rather than
proven for the homogeneous chain: that $\det M\neq0$ for every
eigenstate, and that every transfer matrix eigenvalue is non-degenerate and has a unique corresponding Q-function solving the Baxter equation. In the presence of generic local impurities,
the non-degeneracy and the correspondence between eigenvalues
and Q-functions follow from the operatorial SoV results of
\cite{Kitanine:2018sov,Niccoli:2022xdy}. We expect that
$\det M\neq0$ can likewise be derived by relating our determinant
to the scalar product of separate states of~\cite{Kitanine:2018sov}.
All of these statements are confirmed by our numerical
checks for $L\leq4$, see \secref{sec:generic-explicit-data}. A proof directly for the homogeneous chain, without local impurities, would make the functional
derivation fully self-contained.

There are several possible future directions we now summarise.

\paragraph{Off-diagonal correlators.}
A natural extension concerns matrix elements between
eigenstates of two different transfer matrices, such as
\begin{equation}
\langle\tilde\Psi_A|\sigma^a_\pm|\Psi_B\rangle,
\end{equation}
where $\langle\tilde\Psi_A|$ is an eigenstate of the transfer matrix
with different boundary parameters. The simplest such quantity, the
overlap $\langle\tilde\Psi_A|\Psi_B\rangle$, controls boundary quench
problems in which a boundary field is suddenly changed, and was
recently studied in~\cite{Abetian:2024heu}. Within the modified
algebraic Bethe ansatz, determinant representations for scalar
products of this type were obtained
in~\cite{Belliard:2019ebt,Belliard:2019bfz,Belliard:2021esn},
including the XXZ chain with general integrable boundaries. In the quasi-periodic
case, quantities of this type were computed within Functional SoV
in~\cite{Gromov:2020fwh,Gromov:2022waj}, and the same construction generalises to
rational open spin chains~\cite{Gromov:unpublished}. Combined with the
treatment of the source term developed here, this should give access
to such off-diagonal correlators in the open XXZ chain with generic
boundaries.

\paragraph{More general operator insertions.} The Hellmann--Feynman argument only
reaches operators that arise as parameter derivatives of the
conserved charges, which in the homogeneous chain are boundary
operators. Going beyond these, towards bulk operators and multi-point
correlators, will require functional
analogues of the more general determinant formulae developed for
periodic chains~\cite{Gromov:2022waj}. Since the determinants only
require the Q-function, which can be obtained at large $L$ from the
inhomogeneous Baxter equation, it should also be possible to study
the approach to the thermodynamic boundary magnetisations
of~\cite{Pozsgay:2018ybn} and to extract finite-size corrections at
generic boundaries. In the same spirit, one could take the continuum
limit of the finite-size one-point functions and compare with the
boundary one-point functions of massive integrable quantum field
theories obtained in~\cite{Kormos:2010ae}. More ambitiously, it would be very interesting to
generalise the functional approach so that the SoV integrals can be
evaluated directly in infinite volume, giving access to boundary
observables in the thermodynamic limit without passing through
finite-$L$ determinants.

\paragraph{Higher-rank.}
A natural further direction is to open spin chains of higher rank.
Integrable open chains based on other algebras and their spectra
have been studied extensively, see for
example~\cite{Nepomechie:2017hgw,Frahm:2023lpe,Frahm:2025emg}.
For the $Y(\alg{gl}_n)$ reflection algebra with general integrable
boundary conditions, a SoV basis was constructed
in~\cite{Maillet:2019hdq}, and for $n=2,3$ the transfer matrix
spectrum was characterised by a finite-difference equation of
order $n$ imposed at the separated-variable points, where the
source term vanishes, in analogy with~\cite{Kitanine:2014swa}.
In contrast, the construction presented here uses the full
functional Baxter equation, including the source term.
In the quasi-periodic case, the Functional SoV construction is
already available at any rank, including determinant representations
of scalar products and a complete basis of form
factors~\cite{Cavaglia:2019pow,Gromov:2020fwh,Gromov:2022waj}.
Combining these results with the boundary measure and the treatment
of the source term developed here should give access to boundary
one-point functions for higher-rank open chains with generic
boundaries.

\bigskip

\noindent\hfill\rule{10em}{0.5pt}\hfill\mbox{}

\paragraph{AI Declaration.}

We used Claude Opus 5.5, Sonnet 5.5 (Anthropic) and Wolfbook for
drafting the manuscript and running numerical
checks. The authors are solely responsible for the contents
of this paper.

\paragraph{Acknowledgements.}

This work is partially based on the B.Sc.\ thesis of one of the authors (T.\,Batmunkh). The work of T.\,Bargheer and P.\,Ryan was funded by the Deutsche Forschungsgemeinschaft (DFG, German Research Foundation) -- 460391856.
T.\,Bargheer and P.\,Ryan acknowledge support from DESY (Hamburg, Germany), a member of the Helmholtz Association HGF,
and by the Deutsche Forschungsgemeinschaft
(DFG, German Research Foundation) under Germany's Excellence Strategy
-- EXC 2121 ``Quantum Universe'' -- 390833306.

\bibliographystyle{nb}
\bibliography{references}

\end{document}